\documentclass[article,prd]{revtex4}

\usepackage{amsbsy}
\usepackage{amssymb}
\usepackage{amsmath}
\usepackage{graphicx}

\def\x{{\mathrm{x}}}

\def\n{{\rm n}}
\def\p{{\rm p}}
\def\e{{\rm e}}

\newcommand{\be}{\begin{equation}}
\newcommand{\ee}{\end{equation}}
\newcommand{\beq}{\begin{equation}}
\newcommand{\eeq}{\end{equation}}

\begin{document}

\title{A gravitational-wave perspective on neutron-star seismology}

\author{N. Andersson}

\affiliation{Mathematical Sciences and STAG Research Centre, University of Southampton,
Southampton SO17 1BJ, United Kingdom}

\begin{abstract}
We provide a bird's-eye view of neutron-star seismology, which aims to probe the extreme physics associated with these objects, in the context of gravitational-wave astronomy. Focussing on the fundamental mode of oscillation, which is an efficient gravitational-wave emitter, we consider the seismology aspects of a number of astrophysically relevant scenarios, ranging from transients (like pulsar glitches and magnetar flares), to the dynamics of tides in inspiralling compact binaries and the eventual merged object and instabilities acting in isolated, rapidly rotating, neutron stars. The aim is not to provide a thorough review, but rather to introduce (some of) the key ideas and highlight issues that need further attention.
\end{abstract}

\maketitle

\section{Motivation}

Neutron stars represent many extremes of physics; in density, pressure, temperature (through the early formation stages and during binary mergers), models for which require aspects that may not yet be fully (or perhaps even partially) understood. The composition and state of matter introduce concepts from across modern physics---fluid dynamics and elasticity, thermodynamics, superfluidity and superconductivity, electromagnetism, nuclear physics---while the models have to be developed in the curved spacetime framework of general relativity. To suggest that this is a challenge is not an understatement. In fact, from the theory point of view we may have to accept that we cannot---at least not at this point---account for all aspects we know we ought to consider. Then there are (in the words of a wise philosopher) the ``unknown unknows''... Yet, we want to make progress.

The obvious question to ask is if we can use observations to make sense of the theory mess. This may seem far fetched, given that neutron stars are small and (obviously) distant. They are certainly not hands-on laboratories! Nevertheless, there has been clear progress. Notably, we have precise mass estimates for many systems from radio pulsar timing (telling us, in particular, that the matter equation of state must allow stars with a mass above $2M_\odot$ \cite{2010Natur.467.1081D}) and the recent results from NICER constrain the neutron star radius (very roughly to the range 11-14~km), as well \cite{2019ApJ...887L..21R,2019ApJ...887L..24M}. The constraints on bulk properties---mass and radius---should become increasingly precise as more data become available. Future instruments, like the SKA in the radio and the planned eXTP mission for x-ray timing \cite{2016RvMP...88b1001W}, will ensure that this remains a healthy area of exploration. What is less clear is to what extent this progress will allow us to probe aspects associated with the neutron star interior, e.g. the state and composition of matter.

If we want to explore aspects associated with the dense neutron star interior, it is natural to (try to) formulate a seismology strategy. It is well-known that the complex interior physics is reflected in a rich spectrum of oscillation modes \cite{2019gwa..book.....A} and one may hope to be able to use observations of related features to gain  insight. Similar observation programmes have been been carried out---with enormous success---for both the Sun and distant main-sequence stars. In the case of the Sun, much of the progress was driven by the ESA/NASA SOHO (SOlar and Heliospheric Observatory) space mission. Studies of gravity g-modes and low-multipole pressure p-modes have led to measures of the sound speed at different depths and the differential rotation throughout much of the Sun's interior. For other stars, NASA's Kepler mission has provided key information about the structure and internal dynamics, focussing on oscillations that are stochastically excited by surface convection. This has led to a revolution is the 
field of asteroseismology \cite{2010aste.book.....A}. Interestingly, the same technology allows us to  characterise the host stars of exoplanets, as the inference of  a precise stellar radius constrains the companion planetary radius, as well.

When it comes to neutron stars, we are not likely to be able to ``resolve'' surface features. Instead, it is natural to consider the gravitational-wave aspects of the problem. In essence, any non-axisymmetric deformation/acceleration of the matter in the star will generate gravitational waves and one would expect to be able to express these waves in terms of the star's oscillation modes\footnote{For a Newtonian stellar model, the oscillation modes form a complete set which can be used as a basis to express a general fluid motion. The relativistic problem is more complicated in this respect, but one would nevertheless expect the modes to dominate the star's response to a perturbing agent.}. The promise of such gravitational-wave asteroseismology  \cite{1996PhRvL..77.4134A,1998MNRAS.299.1059A,2001MNRAS.320..307K,2004PhRvD..70l4015B} relies on the answer to two questions. First, are the mode features robust enough that we can use observations to constrain the physics? Second, are there realistic scenarios  where specific oscillation modes are excited to a level that the gravitational-wave signal can be detected by current (or, indeed, future) instruments?
Our aim here is to argue in favour of affirmative answers to both of these questions. This is not to say that this venture will be in any way straightforward, but the effort may pay off handsomely in the end.

\section{The first couple of steps: The f-mode}

As a first illustration of the key aspects, let us consider the simplest setting; a non-rotating uniform density star. 
This problem was  considered by Lord Kelvin back in the late 1800s, although the analysis was carried out in a fashion that would seem arcane to a modern student (as it did not involve spherical harmonics).

In Newtonian gravity,
the dynamics of a non-rotating star is governed by 
the Euler equations (momentum conservation), together with the continuity equation (mass conservation) and the Poisson equation for the gravitational potential. As we are (for now) taking  the density, $\rho$, to be constant, the continuity equation 
simplifies and we have (working in a coordinate basis so that $\nabla_i$ represents the covariant derivative associated with the chosen coordinates)
\be
\partial_t \rho + \nabla_i  (\rho v^i) = 0 \Longrightarrow
\nabla_i  v^i= 0 \ .
\label{cont}\ee
That is, the fluid flow---represented by the velocity $v^i$---is incompressible.
Analysing the problem, we need to distinguish two
kinds of perturbations. Eulerian perturbations relate to changes in the various quantities at a fixed point in space; e.g. $\delta p$ for the pressure (where we obviously have $\delta \rho=0$ in this first example). This is in contrast to co-moving Lagrangian perturbations, which relate to measurements carried out by an observer moving along with the fluid. This  leads to a pressure perturbation $\Delta p$ and if we introduce a displacement vector  $\xi^i$ to connect the perturbed fluid elements
with the corresponding ones in the unperturbed  configuration, we have \cite{1978ApJ...221..937F}
\be
\Delta p = \delta p +  \xi^i  \nabla_i p \ .
\ee
For non-rotating stars, which we focus on for the moment,  the 
displacement vector is simply related to the perturbed velocity
through
\be
\partial_t  \xi^i= \Delta  v^i  = \delta   v^i \ .
\ee
In terms of Eulerian pertubations, the momentum equation takes the form
\be
\partial_t \delta  v_i + 
{ 1\over \rho} \nabla_i \delta p + \nabla_i\delta \Phi = 0 \ ,
\label{peuler}
\ee 
where  $\delta \Phi$ is the variation in the 
gravitational potential, which is governed by
\be
\nabla^2 \delta \Phi = 4\pi \delta \rho \ ,
\ee
such that the right-hand side  vanishes for a constant density model. It follows immediately from these equations that the perturbed velocity must be irrotational and hence it makes sense to introduce a velocity potential such that
\be
\delta v_i = \nabla_i \chi  \ .
\label{vpot}
\ee
The perturbed Euler equations then lead to
\be
\partial_t \chi + { 1\over \rho} \delta p + \delta \Phi = \mbox{ constant} \ .
\label{bern}\ee
It follows that $\chi$, $\delta p$ and $\delta \Phi$ must all solve the homogeneous version of Laplace's equation. This, in turn, tells us that the quantities are naturally expanded in spherical harmonics, $Y_l^m$, and it is easy to show that, in order for the behaviour to be regular at the centre, each quantity must be proportional to $r^l Y^m_l$ for a given $(l,m)$ multipole.
Finally, at the surface of the star, a given oscillation mode must satisfy
\be
\Delta p = \delta p + \xi^r \partial_r p = 0 \qquad \mbox{ at } r=R \ ,
\ee
while the matching to the external gravitational potential requires\footnote{This follows by imposing the relevant junction conditions across the star's surface, accounting the a finite density at $r=R$.}
\be
\partial_r \delta \Phi + { l+1 \over R} \delta \Phi = -4\pi G\rho \xi^r  \ ,
\quad \mbox{ at } r=R \ ,
\ee 
(noting that the right-hand side does not vanish for a constant density model or, indeed, whenever the star's surface is associated with a finite density, as it might be for a quark star).
Working out the algebra (in the Fourier-domain, with a harmonic time dependence $e^{i\omega t}$ for all quantities), we find that
we must have
\be
\omega^2 = { 8 \pi G \rho \over 3} { l(l-1) \over 2l+1 }  \ .
\label{fmode}
\ee
For each value of $l$ we have two  modes (associated with the two signs of the square root), known as the fundamental f-modes.
 The mode frequency scales with the (square root of the) average density of the star and increases with the multipole $l$ (the mode frequencies are the same for all values of $m$ in a non-rotating star). 
The first scaling suggests that it ought to be possible to use observations to constrain the physics of the star. Of course, we also see that the observation of a single mode-frequency would not be enough, as it would only constrain the average density. We  need more information to untangle mass and radius.

Next, in order to estimate the rate at which gravitational-wave emission damps the f-mode we can use the standard post-Newtonian 
multipole formulas \cite{2019gwa..book.....A}. Focussing on a single mode and the contribution from the mass multipole (noting that, in the case of the uniform density model, the contribution comes from the star's surface, where the density is formally discontinuous \cite{1975ApJ...197..203D}) we find the damping timescale 
\begin{equation}
\tau \sim \left({ c^2 R \over GM} 
\right)^{l+1} { R \over c}  \ ,
\label{tgw}\end{equation}
where $M$ is the star's mass, $G$ is Newton's gravitational constant and $c$ is the speed of light.

What do we learn from this? Perhaps the most important insight is this:  A given uniform density model is described by two parameters, which we can take to be the density and the radius. An observation of the mode frequency would provide the former and then the damping rate would help us infer the radius---as the scaling with the parameters is different. This is the main idea of gravitational-wave asteroseismology  \cite{1996PhRvL..77.4134A,1998MNRAS.299.1059A,2001MNRAS.320..307K,2004PhRvD..70l4015B} and it prompts a number of follow-on questions. 

How likely are we to be able to infer both the frequency and the damping rate from observations? The answer depends on the energy associated with the mode excitation, but it is generally clear that---since these modes are rapidly damped---the oscillation frequency is easier to pin down than the damping. As an illustration, consider the example from \cite{2019gwa..book.....A} (which draws on the empirical relations from \cite{2001MNRAS.320..307K}) which shows that---for a typical f-mode with frequency $f=2.4$~kHz and damping time $\tau=0.1$~s,  a galactic source at a distance of $10$~kpc and a fiducial detector with sensitivity (spectral noise density) $S_n^{1/2} = 10^{-23}$~Hz$^{-1/2}$ at the mode frequency (roughly the level  of advanced LIGO instruments during the O3 science run),  we expect a signal-to-noise ratio (not to be confused with the matter density)
\be
\rho_\mathrm{n} \approx 30 \left( {E \over 10^{-6} M_\odot c^2} \right)^{1/2} \ , 
\ee
where $E$ represents the energy channeled through the specific mode we consider\footnote{Here and in the following we adopt the relativist's convention of relating to energies in terms of the solar mass equivalent. Anyone of an astrophysics persuasion may want to keep in mind that $M_\odot c^2 \approx 2\times 10^{54}$~erg.  }. We see that, if the system radiates about $E=4\times 10^{-7}M_\odot c^2$ through this mode we would detect the signal with $\rho_\mathrm{n}\approx 10$. 
The relative error involved in the parameter estimation may be estimated by a Fischer matrix analysis \cite{2001MNRAS.320..307K}. In the suggested example, this leads to
\be
{\delta f / f} \approx 1.3\times10^{-3}/\rho_\mathrm{n} \ , 
\ee
and
\begin{equation}
{\delta \tau / \tau}\approx 2/\rho_\mathrm{n} \ .
\ee
For a signal leading to $\rho_\mathrm{n} \approx 10$ we would accurately infer the mode frequency but the damping rate would only be known at the 20\% level.  If we want to extract both the frequency and the damping rate to (say) the 1\% level, then we need a signal-to-noise ratio of at least 200. This would require the release of an energy of order $4\times10^{-5}M_\odot c^2$, which---as we will discuss below---may be unrealistic. 

This brings us to the astrophysics. Are there realistic scenarios where different oscillation modes are excited to a significant amplitude? We will consider some possibilities later. And finally, what about more realistic stellar models? Are the suggested scaling relations robust for a range of (say) compressible equations of state? On the one hand, we want this to be the case as the observable features should then be robust, as well. On the other hand, we need there to be differences, as the main aim of the exercise is to identify the underlying matter properties and this (obviously) means that we need to be able to tell the difference between different models. 

The last point suggests that we need to consider more realistic implementations based on  relativistic neutron star models. Naturally, this calculation is a bit more involved. First of all, we need to account for the compressibility of matter. This means that we have to generalise the form for the displacement vector. Instead of working with the gradient of a scalar potential, we now have
\beq
\xi^i =  {W(r) \over r }\nabla^i r  + V(r) \nabla^i Y_{lm} \ , 
\label{displace}
\eeq
We also need to consider the density variations. These are linked to the pressure via the equation of state, and depend explicitly on the speed of sound, $c_s$, in the stellar fluid. For a barotropic model with  $p=p(\rho)$, we have
\be
c_s^2 = {d p \over d \rho} \quad \mbox{and}\quad   \Delta p = {p\Gamma_1 \over \rho} \Delta \rho = c_s^2 \Delta \rho   \ .
\ee
Notably, the background equilibrium and the Lagrangian perturbations are described by ``the same'' equation of state in this case. As we will see later, this  is not generally the case.

Working through the details, we now find (in addition to the f-mode) a set of higher frequency pressure p-modes. These depend directly on the compressibility of matter. Higher overtone p-modes have---for a given multipole $l$---increasingly high frequecies. The p-modes tend to be less efficient emitters of gravitational waves as they have radial nodes in the eigenfunctions which lead to cancellations in the mass multipole integral \cite{1995MNRAS.274.1039A}. Moreover, as the typical p-mode frequency lies above several kHz, these modes will be difficult to detect with ground-based instruments (as these tend to be designed with binary inspirals in mind, leading to a sensitivity sweet spot around 100~Hz).

What happens when we consider the problem in general relativity, as we know we must if we want to use a realistic matter description? 
Qualitatively, the problem is different because the modes become ``quasi''-normal. In a live spacetime, they are determined from an asymptotic condition of purely outgoing gravitational waves. This means that they have complex frequencies, with the imaginary part representing the damping rate---we no longer have to resort to the multipole formula, but the mode solution is more complicated as the outgoing wave condition needs to be implemented with care \cite{1995MNRAS.274.1039A,1993PhRvD..48.3467L,2003CQGra..20.3441G}. Fortunately, there are a number of robust methods for dealing with this issue.

Focussing on the f-mode, the relativistic results largely bring out the expectations from the Newtonian calculation. The results from \cite{1996PhRvL..77.4134A,1998MNRAS.299.1059A} demonstrate that the scalings from \eqref{fmode} and \eqref{tgw} work pretty well, leading to useful phenomenological relations for the frequency and the damping rate. This is not too surprising because, even though the microphysics associated with different equations of state may be different, the neutron star bulk properties are fairly robust. In fact, Lattimer and Prakash have pointed out \cite{2001ApJ...550..426L} that the old 
Tolman VII solution to the Einstein equations \cite{1939PhRv...55..364T}, which takes as the starting point a quadratic density distribution,
accords extremely well with the results for a wide range of ``realistic'' equations of state. This, in turn, suggests that we should not be too surprised if different neutron star properties turn out to be related by more or less ``universal'' relations \cite{2013Sci...341..365Y}. 

Returning to the f-mode, the relativistic mode frequencies encode a dependence on the gravitational redshift---as the modes are ``observed'' at infinity. This suggests an additional dependence on the star's compactness, $M/R$, and hence a more complicated parameter scaling. The results from \cite{2005MNRAS.357.1029T} show that a scaling in powers of the stellar compactness leads to a useful f-mode relation for a range of matter models. An even tighter relation is provided in \cite{2010ApJ...714.1234L}, based on working with $\eta =(M^3/I)^{1/2}$, where $I$ is the star's moment of inertia, as the scaling parameter. These relations are accurate enough that a seismology strategy can be formulated for realistic neutron star models. 

\section{Transients}

For isolated neutron stars, one might envisage a number of scenarios that lead to transient gravitational-wave emission which may, in turn, be channeled through individal modes of oscillations. In fact, if the modes form a complete set (in the mathematical sense) then it should be possible to represent any given ``initial data'' as a mode-sum. This should be the case in Newtonian theory (although the problem is intricate for rotating stars \cite{1978ApJ...222..281F}), but the modes are not expected to be complete in relativity\footnote{Essentially, the inevitable gravitational-wave damping makes the problem non-Hermitian, and the evolution problem also involves a late-time power-law tail associated with backscattering of waves by the spacetime curvature \cite{1994PhRvD..49..883G}.}.  In essense, we consider the likely energy budget for different ``explosive'' phenomena and then, through a back-of-the-envelope argument,  estimate the gravitational-wave amplitude. This is instructive, but the exercise comes with important caveats. In particular, an energy based argument that something is possible in principle does not in any way demonstrate that this happens in reality---a simple reflection of the fact that we need the energy to be associated  with asymmetries in the fluid motion in order to lead to gravitational-wave emission (a spherical expansion/contraction simply won't do!). The converse is, of course, true. If we can show that there is not enough energy  available to excite a given mode to a relevant amplitude then the proposed scenario is unlikely to be of interest.

We can estimate the gravitational-wave strain $h$ from the standard flux formula \cite{2019gwa..book.....A}
\begin{equation}
{ c^3 \over 16 \pi G} |\dot{h}|^2 = { 1 \over 4 \pi d^2}  \dot E \ , 
\label{flux1}
\end{equation}
where  $d$ is the distance to the source, and the dots represent time derivatives. In our case, we characterize a given event by the damping time
$\tau$ and assume that the signal is monochromatic, with mode-frequency $f$. Then we  use $\dot E\approx E/\tau$ and $\dot{h} \approx 2 \pi f h$ to get 
\begin{equation}
h \approx 4\times 10^{-23} \left( { E \over 10^{-9} M_\odot c^2} \right)^{ 1/2}
\left({ \tau \over 0.1 \mbox{ s} }\right)^{-1/2}
\left( {f\over 2 \mbox{ kHz}} \right)^{-1} \left( { d\over 10 \mbox{ kpc}}\right)^{-1} \ ,
\label{hraw}\end{equation}
This rough estimate accords fairly well with the more precise statement of detectability from before.  The key point is that, for the  energy budget to be reasonable we have to focus on galactic events.

What kind of---ideally regularly occurring---astrophysical event may lead to a neutron star exhibiting ``large'' amplitude oscillations? The traditional example would be the  supernova explosion in which the star is born. However,  estimates for the energy radiated
as gravitational waves from supernovae tend to be rather pessimistic,
suggesting a total release of no more than the equivalent to $10^{-6} M_\odot
c^2$, or so. This might still be enough to secure a detection, but as we only expect a few galactic supernovae per century we may need a fair bit of patience. Nevertheless, recent work has demonstrated that there are interesting seismology aspects to the core collapse  problem \cite{2006NewAR..50..487B,2019ApJ...876L...9R}, so it is clearly worth pursuing this in more detail. 

Another potential excitation mechanism
for stellar oscillations would be some kind of starquake, e.g., associated with a pulsar
glitch or a magnetar flare. The idea is that the evolution of the star---e.g. associated with magnetic field decay--- leads to strain in the elastic crust and that the stored energy is suddenly released when the system reaches a critical level. The typical energy released in this process would then be of the
order of the maximum mechanical energy that can be
stored in the crust, estimated to be at the level of $10^{-9}-10^{-7}M_\odot
c^2$ \cite{1989ApJ...343..839B,1998ApJ...500..374M}. This suggestion is interesting given that magnetars appear to be associated with fairly regular flaring events \cite{2016RvMP...88b1001W}. If modes are excited
in these systems, an indication of the energy released in the most powerful
bursts is the  $10^{-9}M_\odot
c^2$ estimated for the March 5 1979 burst in SGR~0526-66 \cite{1983A&A...126..400B}. However, if the main action leading to the energy emission is associated with the low-density crust one would not expect significant gravitational-wave emission. We may use observations as basis for an asteroseismology analysis---as has indeed be done \cite{2007MNRAS.374..256S}---but the gravitational-wave aspect would be missing. This does not means that we should not search for counterpart signals to the observed flares---as has, indeed, been done \cite{2005PhRvL..95h1103B,2007PhRvD..76f2003A,2008PhRvL.101u1102A,2009ApJ...701L..68A,2011ApJ...734L..35A}---but we should do this with realistic expectations.

The situation is somewhat similar for pulsar glitches, although these events involve significantly less energy. In this case the release of energy is assumed to be associated with large-scale superfluid vortex unpinning in the outer core/inner crust of the star \cite{2015IJMPD..2430008H}. A simple energy estimate suggests that these events may be interesting for the dedicated gravitational-wave astronomer \cite{2001PhRvL..87x1101A}, but a closer analysis indicates that the excitation of the large-scale fluid motion we require may be less likely \cite{2010MNRAS.405.1061S}. Still, searches for glitch-related transients in LIGO gravitational-wave data have already been carried out \cite{2011PhRvD..83d2001A}.

Yet another scenario associated with the star's evolution relates to internal phase transitions. The idea is simple. As the star spins down---due to the expected electromagnetic braking torque---the centrifugal force decreases and the star's central density increases. If the equation of state predicts a sharp phase transition at some density, e.g. linked to  quark deconfinement, then the star might rapidly contract when this density is reached. This contraction leads to potential energy being converted into heat and, perhaps, gravitational waves, as well. The latter component may, however, be small unless one can think of a way for the phase transition to induce asymmetries. At the end of the day, the different evolutionary scenarios are perfectly reasonable, but---in each case---it is difficult to predict the actual level of gravitational-wave emission. From the theory point of view this is problematic, but  
 this does not mean that observers should not look for this kind of signal. Perhaps we need an observational breakthrough to get the theory right.

Another issue with transient events is that one would have to distinguish an astrophysical signal from instrumental artefacts; something that goes ``ping'' in the detector. Both may look like exponentially damped sinusoids, so how do we tell the difference? In this respect, it is interesting to note the transient candidate S191110af, announced by the LIGO/Virgo collaboration on November 10 2019. The suggested ``signal''  consisted of a 1.78~kHz oscillation lasting 0.104~s
\cite{chatterjee19}. 
Follow-up analysis identified instrumental
artefacts in the data, which led to retraction of S191110af as a
genuine signal \cite{chatterjee19b}. However, in the intervening time it was noted that the parameters were consistent with the f-mode of a fairly typical neutron star, see  \cite{kaplan} and  the simple estimates from above. It was also noted that there was no  evidence that a  glitching pulsar  could be responsible for
S191110af \cite{kaplan}. The incidence is nevertheless interesting. The key point is that, while the particular event is unlikely to have been of astrophysical origin,  we have seen that neutron stars can  exhibit transients
at the required  level. Hence, it is natural to ask if we can design a strategy that makes a distinction between astrophysical transients and instrumental features. An interesting suggestion in this direction is outlined in \cite{2020PhRvD.101j3009H}. 
The idea is to bring in additional information. First of all, a transient associated with (say) an f-mode oscillation would have to correlate with both the frequency and the damping rate. However, we know from the example of S191110af that this may not be enough. In the ideal case we would have an electromagnetic counterpart signal---a gamma-ray flare, an x-ray feature or perhaps a radio pulsar glitch---but one might have to be very lucky to make such an observation. Another reasonable strategy would be to draw on the known neutron star population. Is there for example an excess of transients with ``neutron star-like'' features and an energy distribution such that is could reasonably be associated with the galactic neutron star population? The argument then becomes statistical rather than based on a single outstanding event. This raises other issues, but it is one direction in which we might be able to make progress. 

\section{Adding physics: The g-modes}

So far we have not considered the complex physics that characterises the neutron star interior. Additional features come into play when we consider the matter description beyond the pressure-density relation; when we account for the composition and state of matter. As a simple rule-of-thumb,  each additional  piece of ``physics'' brings (at least) one new class of oscillation modes into existence. This makes the seismology problem richer, but also more complicated as we need to keep track of new parameters and these may not be well constrained by our understanding of the nuclear physics. The pessimist may see an insurmountable obstacle here, but the optimist cannot help being excited by the discovery potential.

If we add internal stratification associated with, for example, 
temperature or composition gradients (like a varying proton fraction), then the so-called gravity
g-modes come into play \cite{1992ApJ...395..240R}. One immediate way of seeing this is to 
 introduce the adiabatic index of the perturbations,
$\Gamma_1$, as before; 
\be
    {\Delta p \over p} = \Gamma_1 {\Delta \rho \over \rho} \ ,
\ee
leading to  (for a spherical background star)
\be
    \delta p = {p \Gamma_1 \over \rho} \delta \rho +  \xi^i 
               \left[{p \Gamma_1 \over \rho} \nabla_i \rho - \nabla_i p 
               \right]\equiv {p \Gamma_1 \over \rho} \delta \rho + 
               p \Gamma_1 ({ \xi^r} A) \ ,
\label{schw}
\ee
which defines the  Schwarzschild discriminant $A$. This adds a restoring force (buoyancy) for the perturbed fluid and leads to the presence of g-modes. 

It is crucial to consider the impact of nuclear reactions on the composition g-modes. Following \cite{2019MNRAS.489.4043A} we take as our starting point a two-parameter equation of state $p=p(\rho,x_\p)$, where $x_\p=n_\p/n$ is the proton fraction (in terms of the proton and baryon number densities, $n_\p$ and $n$), and work in the context of Newtonian gravity. Then the mass density is simply $\rho = m_B n$ where $m_B$ is the baryon mass. In essence, we can think of $\rho$ as a proxy for the baryon number density, which is conserved (by virtue of the continuity equation).
In order to account for nuclear reactions, we  introduce a new dependent variable $\beta=\mu_\n-\mu_\p-\mu_\e$ which encodes the deviation from chemical equilibrium (with $\mu_\x$ and $\x=\n,\p,\e$ representing the chemical potential for neutrons, protons and electrons, respectively). For simplicity, we assume a pure npe-matter neutron star core, which means that the relevant reaction timescales are those associated with the Urca reactions. Formally, we then have the thermodynamical relation (as we focus on the matter composition, we ignore the temperature here)
\beq
dp = \sum_\x n_\x d\mu_\x = n ( d\mu_\n - x_\p d\beta) \ ,
\label{dprel1}
\eeq
where we have assumed (local) charge neutrality ($n_\p=n_\e$). As the background configuration is in both hydrostatic and beta equilibrium we only account for reactions at the level of the perturbations. We then, first of all, note that \eqref{dprel1} should also hold  for Lagrangian perturbations, so we have
\beq
\Delta p =  n ( \Delta\mu_\n - x_\p \Delta\beta)  \ ,
\eeq
or, as it turns out to be more convenient to work with $\rho$ and $\beta$,
\be
\Delta p = \left( {\partial p \over \partial \rho} \right)_{\beta} \Delta \rho +  \left( {\partial p \over \partial \beta} \right)_{\rho} \Delta \beta 
= c_s^2  \Delta \rho +  \left( {\partial p \over \partial \beta} \right)_{\rho} \Delta \beta \ ,
\label{eqone}
\ee
where $c_s$ is sound speed for the background equilibrium.

How do we account for nuclear reactions? Well, 
for the protons we then have 
\beq
 (\partial_t + v^j \nabla_j ) n_\p +n_\p \nabla_j v^j  =  \Gamma \ ,
\eeq
with $\Gamma$ the relevant reaction rate (not to be confused with the adiabatic index $\Gamma_1$). Combining this with overall baryon number conservation (assuming that all components move together!) and assuming that the reaction rate relates to perturbations, we have 
\beq
\Delta \left[  (\partial_t + v^j \nabla_j ) x_\p \right] =  (\partial_t + v^j \nabla_j )\Delta x_\p  = {\Gamma\over n}  \ , 
\label{eqtwo}
\eeq
where, at least for small deviations from equilibrium  \cite{1992A&A...262..131H,1995ApJ...442..749R},
\beq
\Gamma \approx \gamma \Delta \beta \ ,
\eeq
with   $\gamma$ encoding the relevant reaction rate.

Now thinking of $\beta$ as a function of $\rho$ and $x_\p$, and assuming that the star is non-rotating (so that $v^i=0$), we  have
\beq
\partial_t \Delta \beta = \left( {\partial \beta \over \partial \rho}\right)_{x_\p}  \partial_t \Delta \rho +  \left( {\partial \beta \over \partial x_\p}\right)_{\rho} \partial_t  \Delta x_\p 
 = \left( {\partial \beta \over \partial \rho}\right)_{x_\p}  \partial_t \Delta \rho +  \left( {\partial \beta \over \partial x_\p}\right)_{\rho} {\gamma \over n} \Delta \beta \ . 
\eeq
That is, we have
\beq
\partial_t \Delta \beta -  \mathcal A  \Delta \beta = \mathcal B \partial_t \Delta \rho \ ,
\label{eqfour}
\eeq
with
\beq
\mathcal A =  \left( {\partial \beta \over \partial x_\p}\right)_{\rho} {\gamma \over n} \ , \quad \mbox{and} \quad
\mathcal B = \left( {\partial \beta \over \partial \rho}\right)_{x_\p} \ .
\eeq

If we work in the frequency domain (essentially assuming a time-dependence $e^{i\omega t}$ for the perturbations, and taking the coefficients $\mathcal A$ and $\mathcal B$ to be time independent)  we have
\beq
\Delta \beta = {\mathcal B \over 1 + i \mathcal A/\omega} \Delta \rho \ .
\label{dbeta1}
\eeq
At this point we can consider the timescales involved in the problem. Introducing a characteristic reaction time as 
\beq
t_R = {1 \over \mathcal A} \ ,
\eeq
(noting that the actual timescale is the absolute value of this)
we see that, if the reactions are fast compared to the dynamics (associated with a timescale $\sim 1/\omega$) then $|t_R \omega| \ll 1$ and it follows that
\beq
\Delta \beta \approx 0 \ .
\eeq
Basically, because the dynamics are slow compared to the reactions the fluid remains in beta equilibrium. There will be no buoyancy and hence no g-modes. The situation changes in the limit of slow reactions, where  $|t_R \omega| \gg 1$ and we can Taylor expand \eqref{dbeta1} to get
\beq
\Delta \beta \approx \mathcal B \left(  1-  i \mathcal A/\omega \right)  \Delta \rho \approx \mathcal B \Delta \rho \ .
\label{dbeta}
\eeq
Using this result in \eqref{eqone}, we have
\beq
\Delta p = \left[ \left( {\partial p \over \partial \rho} \right)_{\beta} + \left( {\partial p \over \partial \beta} \right)_{\rho} \left( {\partial \beta \over \partial \rho}\right)_{x_\p} \right]  \Delta \rho 
\equiv \mathcal C \Delta \rho \ , 
\label{eqfive}
\eeq
which leads to
\beq
\delta p = \mathcal C \delta \rho + \left[ \mathcal C \xi^j \nabla_j \rho - \xi^j \nabla_j p \right]\ .
\eeq
with $\mathcal C = p\Gamma_1/\rho$.
However, since
\beq
\nabla_j p = \left( {\partial p \over \partial \rho} \right)_{\beta} \nabla_j \rho = c_s^2 \nabla_j \rho\ ,
\eeq
we are left with
\beq
\delta p = \mathcal C \delta \rho +  \left( {\partial p \over \partial \beta} \right)_{\rho} \left( {\partial \beta \over \partial \rho} \right)_{x_\p}  \xi^j \nabla_j \rho \ .
\label{dprel}
\eeq
In this case, the composition of matter affects  both terms on the right-hand side of the relation. The mathematics may be fairly straightforward, but the physics is less so. In particular, we now need an equation of state that allows us to calculate the different thermodynamical derivatives used to relate \eqref{dprel} and \eqref{schw}. While there are several such models ``on the market'' (a notable example being the results from the Brussels-Montreal collaboration \cite{2011PhRvC..84f2802C}) it is clear that we are asking questions that go beyond the  description of equilibrium configurations---we are beginning to touch upon issues like the relevant transport properties associated with the matter \cite{2018ASSL..457..455S}.

This, fairly lengthy, argument may seem like a detour but it is important as it helps us understand a number of issues. First of all, we see that it is straightforward to analyse the problem in the two limits of fast and slow reactions. In the intermediate regime, where the dynamical timescale is similar to that of reactions we have to explicitly solve for the evolving proton fraction. Secondly, we see that the presence of g-modes depend entirely on the timescales. For example, the relatively slow Urca reactions (we will quantify this later) will allow g-modes associated with a varying proton fraction \cite{1992ApJ...395..240R}, but  as the relevant reactions are much faster there may be no g-modes arising from a varying hyperon fraction. Detailed mode calculations---ignoring the argument we have just outlined---suggest that the g-mode spectrum has a infinite number of ``overtones'' with decreasing frequency. Our argument obviously impacts on this. If we account for the reaction rates, then no matter how slow they are, there has to be a point where the oscillation of a (perhaps very) high order g-mode is even slower. This mode should not exist, at least not as an oscillation feature \cite{2019MNRAS.489.4043A}. In essence, the spectrum of composition g-modes will actually be finite for a realistic neutron star model. The may seem like a fine-print issue, but it could be conceptually important (as in the tidal problem we consider below).

It is easy to see that thermal gradients may also lead to the presence of g-modes. The previous logic applies. We need the dynamics to be fast compared to the thermal evolution. Thermally supported g-modes are expected to be relevant for proto-neutron stars \cite{2003MNRAS.342..629F,2004PhRvD..70h4009G}.
The state of matter also affects the g-modes. For example, in a superfluid neutron star core---which requires a two-fluid model \cite{2021Univ....7...17A}---the varying proton fraction will not introduce buyoancy. The g-modes disappear \cite{1995A&A...303..515L,2001MNRAS.328.1129A}, at least until we reach the density where muons enter the game. At this point there will be a composition gradient, now associated with variations in the muon fraction. This re-introduces the g-modes, but at a slightly higher frequency (roughly a factor $1/x_\p$ higher than in the case of npe-matter) \cite{2013PhRvD..88j1302G,2016MNRAS.455.1489P}.

\section{Dynamical tides}

When we outlined  scenarios for transient mode excitation we considered different mechanisms internal to the star. One may also envisage oscillations being driven by an external agent. The natural example of this is the tidal interaction in a binary system. When the two components of a binary system are far apart, a point-particle description of each body is sufficient to describe the orbital motion. This assumption is adequate for both classic binary dynamics and the post-Newtonian approximation used to model gravitational-wave signals from compact binaries \cite{2006LRR.....9....4B}. However, the situation changes as the stars are brought closer together through the emission of gravitational waves.
Finite size effects come into play during the late stages of inspiral, with the tidal deformability \cite{2008PhRvD..77b1502F,2008ApJ...677.1216H,2010PhRvD..81l3016H} of the supranuclear density matter, 
\beq
\Lambda_l =  {2\over (2l-1)!!} {k_l \over \mathcal C^{2l+1}} \ , 
\label{lamdef}
\eeq
where $k_l$ is the Love number for the $l$-multipole and $\mathcal C=M/R$ is the stellar compactness,
leaving an imprint on the gravitational-wave signal. As  demonstrated in the celebrated case of  GW170817---the first observed neutron star binary merger \cite{2017ApJ...848L..12A,2018arXiv180511581T}---this leads to a constraint on the neutron star radius \cite{2018PhRvL.121p1101A,2018PhRvL.121i1102D} and hence the equation of state. 

As it turns out, a precise understanding of the tidal response requires an analysis of both the state and composition of matter. One aspect of this relates to the composition of matter, which is likely to remain ``frozen'' during the late stages of binary inspiral. It is easy to see that this should be the case.
A typical neutron star binary system may spend  10-15 minutes in the sensitivity band of a ground-based interferometer (above 10 Hz). 
As the stars are deformed by the tidal interaction, matter is  driven out of equilibrium but the relevant nuclear reactions are (likely to be) too slow to re-establish equilibrium on this timescale. This is evident from, for example, the estimates in \cite{1992A&A...262..131H}, which suggest that the  equilibration timescale for the direct Urca reactions is 
\beq
t_D \sim 0.1~\mathrm{month}  \left({T\over 10^8~\mathrm{K}} \right)^{-4} \ , 
\eeq
with the modified Urca relaxation a factor of about $10^7$ slower (at $10^8$~K). Given that inspiralling binaries are mature---and hence cold---we can safely assume that the equilibration time will be much longer than the time it takes a given system to move through the sensitivity band of a ground-based interferometer (months-years vs minutes). In fact, at the expected core temperature the star's interior should be superfluid, in which case reactions  are suppressed (exponentially), see \cite{2017MNRAS.464.2622Y} for comments on how superfluidity impacts on the tidal problem (a problem with a number of interesting aspects). In essence, the  equilibration argument supports the assumption that  the equation of state is no longer barotropic, as has (indeed) been assumed in virtually every analysis of the tidal problem (see for example \cite{1994MNRAS.270..611L,1994ApJ...426..688R,1995MNRAS.275..301K,1999MNRAS.308..153H}). 

There are two main aspects to the tidal problem. First of all, the tidal deformability, usually represented by the Love numbers \cite{2008PhRvD..77b1502F,2008ApJ...677.1216H,2010PhRvD..81l3016H}, provide the static response during an adiabatic inspiral. Secondly, individual oscillation modes may become resonant with the tidal driving, leading to a distinct contribution for some specific frequency range. The two aspects may be explored separately (e.g. static deformations vs time-dependent ones) but as they are related it is useful to consider them within the same framework. 

In the tidal case, the perturbed Euler equation (for a non-rotating and compressible star) is
\beq
\partial_t^2  \xi_i +{1\over \rho}  \nabla_i \delta p - {1\over \rho^2} \delta \rho \nabla_i p + \nabla_i \delta \Phi = - \nabla_i \chi
\label{eul2}
\eeq
alongside the usual continuity equation and the Poisson equation for the gravitational potential. We also need
 the tidal potential $\chi$ (due to the presence of the binary partner), which is given by a solution to $\nabla^2 \chi = 0$.
 In  a coordinate system centred on the primary, which we will take to have mass $M$, we have \cite{1999MNRAS.308..153H}
\beq
\chi =  - GM'\sum_{l\ge2} \sum_{m=-l}^l {W_{lm} r^l \over D^{l+1}(t) } Y_{lm} e^{-im\psi(t)} \ , 
\label{tidpot}
\eeq
where $M'$ is the mass of the secondary and $D$ is the orbital separation. The orbit of the companion is taken to be in the equatorial plane and $\psi$ is the orbital phase. For $l=2$ (which dominates the gravitational-wave contribution) we have
\beq
W_{20}= - \sqrt{\pi/5} \label{W20} \ , \quad W_{2\pm1} = 0 \ , \quad
W_{2\pm2} = \sqrt{3\pi/10}
 \ .
\eeq

We now want to  express  the driven response of the stellar fluid in terms of the (presumably complete\footnote{Mathematical completeness is not an absolute requirement, but in order for a mode-sum representation to be useful it must be the case that the modes we include dominate the response. In the tidal problem, the (by far) largest contribution is made by the f-mode (at least for non-rotating stars) \cite{1994ApJ...432..296R,2020PhRvD.101h3001A} so we do not have to worry too much about the formal aspects here.} ) set of normal modes. These correspond to solutions $\xi_n$ (where $n$ labels  the different modes in a suitable way). Working in the frequency domain, i.e. solving the Fourier transform version of \eqref{eul2}, and letting the  mode-frequency be $\omega_n$ (real-valued in absence of dissipation) we have the mode-sum:
\beq
 \xi^i = \sum_n a_n   \xi_n^i \ , 
\eeq
where each individual mode satisfies
\beq
-\omega_n^2 A   \xi_n^i  + C  \xi_n^i=0 \ .
\label{fseq}
\eeq
Here $A=\rho$, while the $C$ operator is messy but we do not need the explicit expression in the following. We will, however, make use of the orthogonality of the (time-independent) eigenfunctions. This is demonstrated by introducing the inner product from \cite{1978ApJ...222..281F}. For two solutions $\eta^i$ and $\xi^i$ to \eqref{fseq}, we define
\be
\langle \eta^i,\xi_i \rangle = \int (\eta^i)^* \xi_i dV \ ,
\ee
where the asterisk denotes complex conjugation. It follows that (suppressing the component indices to avoid cluttering up the expressions)
\be
\left<\eta, A\xi\right>=\left<\xi,A\eta\right>^* \ \ ,
\ee
and
\be
\left<\eta, C\xi\right> = \left<\xi,C\eta\right>^* \ .
\ee
From these symmetry relations it is relatively easy to show that two mode solutions $\xi^i_n$ and $\xi^i_{n'}$ are orthogonal as long as $n\neq n'$. That is, we have (keeping the normalisation of the modes explicit)
\beq
\langle \xi_{n'} , \rho \xi_n \rangle = \mathcal A_n^2 \delta_{n n'} \ ,
\eeq
where  \eqref{displace} leads to (for polar modes, like the f-mode)
\beq
\mathcal A_n^2 = \int_0^R \rho \left[ W_n^2 + l(l+1) V_n^2\right] dr \ .
\eeq

We  can also use the  orthogonality to rewrite \eqref{eul2} as an  equation for the mode amplitudes:
\beq
\ddot a_n + \omega_n^2 a_n = - {1\over \mathcal A_n^2} \langle \xi_n, \rho \nabla \chi\rangle \ . 
\eeq
Making use of the  continuity equation
and integrating by parts we have (assuming that the density vanishes at the surface of the star)
\beq
- \langle \xi_n, \rho \nabla\chi\rangle = - \int \rho (\xi^i_n)^* \nabla_i \chi dV =  \int \chi  \nabla_i (\rho \xi^i_n)^* dV= - \int \chi \delta \rho^*_n dV \ . 
\eeq
Given the harmonic time dependence associated with the Fourier domain ($\sim e^{i\omega t}$) we have an equation for the amplitude of tidally driven modes (for each $l$, as the different values of $m$ are  degenerate for non-rotating stars):
\beq
a_n =   {1 \over \omega_n^2 - \omega^2}  {Q_{n} \over \mathcal A_n^2}  v_{l} \ , 
\label{reason}
\eeq
where $v_l$ follows from \eqref{tidpot} (after integrating out the angles) and
 we have introduced the ``overlap integral''
\beq
Q_{n} = - \int \delta \rho^*_n r^{l+2}  dr \ . 
\label{overlap}
\eeq
Noting that the frequency support of the tidal potential (for slowly evolving orbits) leads to $\omega = m\Omega$, where $\Omega$ is the orbital frequency, we have a driven set of modes
which may become resonant during a binary inspiral  \cite{1994MNRAS.270..611L,1994ApJ...426..688R,1995MNRAS.275..301K}.

The Poisson equation allows us to relate the density perturbation to the gravitational potential. The matching at the star's surface then relates the result to the relevant multipole moments, and we have
\beq
Q_{n}  =   { 2l+1 \over 4\pi G } R^{l+1}  \delta  \Phi_n (R) =  I_n \ ,
\eeq
where $I_n$ is the contribution each mode makes to the mass multipole moment. We can then quantify, for example, how much energy is deposited in a given mode during an inspiral. We can also express the result for the dynamical tide as an effective tidal deformability/Love number. The result we want follows immediately from 
\beq
k_l = {1\over 2} {\delta \Phi(R) \over \chi(R) } \ .
\eeq
Expressed as a sum over the modes of the star \cite{2020PhRvD.101h3001A}, we have
\beq
k_l^\mathrm{eff} =  -{1\over 2} + {1\over 2 R^l} \sum_n {Q_n \over \mathcal A_n^2} {1 \over \omega_n^2 - (m\Omega)^2}\left[ \omega^2 V_n(R) - {GM\over R^3} W_n(R)\right] \ .
\label{keff1}
\eeq
This provides us with a useful expression for the dynamical tide, calculable once we solve for star's oscillation modes. Basically, the tidal response is a seismology problem \cite{2020NatCo..11.2553P,2021MNRAS.tmp..395A}.

In addition to  the dynamical tide, the mode-sum  provides a representation for the static tide. In the low-frequency driving limit ($\Omega \to 0$) we retain the usual Love number. The relation is exact for the constant density model, where we only have to consider the f-mode. We then get \cite{2020PhRvD.101h3001A}
\beq
k_l^\mathrm{eff} =  { \omega_n^2 - GMl/R^3  \over  2 \left[(m  \Omega)^2 - \omega_n^2\right]} \to {3\over 4 (l-1)} = k_l \quad \mbox{for}\ m \Omega  \ll  \omega_n \ .
\label{klimit}
\eeq
This is the expected result for the Love number \cite{2014grav.book.....P}.

In general,  the mode-sum allows us to quantify the level at which the matter composition enters the problem. Evidence for simple model problems (with a fixed value for $\Gamma_1$ providing the stratification) shows that the f-modes vastly dominate the tidal response, with p- and g-modes contributing at (most at) the few percent level \cite{2020PhRvD.101h3001A}. This is an interesting observation given that third generation (3G) gravitational-wave detectors like the Einstein Telescope or the Cosmic Explorer may be able to constrain the tidal deformability to the few percent level \cite{2016PhRvD..93k2004M}. It would certainly seem warranted to ask if this could allow us to use observations to constrain (some aspects of) the internal composition.

The mode sum allows us to make progress on a number of issues. For example, assuming that the f-mode dominates the mode sum---which seems a safe bet---it is easy to understand why we should expect to find a robust ``universal relation'' between the mode frequency and the tidal deformability. This result was  demonstrated in \cite{2014PhRvD..90l4023C} and recent evidence suggests that the relation remains accurate for a wide range of equation of state models. Given this observation, we can  combine the mode sum with the phenomenological relation from \cite{2014PhRvD..90l4023C} to write down---adding assumptions about the red shift of the  frequencies involved---a similar phenomenological expression for the dynamical tide. As outlined in \cite{2021MNRAS.tmp..395A}, this leads to
\beq
k_l^\mathrm{eff} \approx {\bar \omega_f^2 k_l \over \bar \omega_f^2 - \delta (2\bar \Omega)^2} +{(2\bar \Omega)^2 \over \bar \omega_f^2 - \delta (2\bar \Omega)^2}\left[ {\delta \over 2} - {\bar \omega_f^2 \over \mathcal C^3 } {\epsilon\over l} \left( k_l + {1\over 2} \right) \right]\ ,
\label{kleff3}
\eeq
where $\bar \omega_f=\omega_f M$ is the f-mode frequency ( the scaling with the star's mass  leading to the natural dimensionless combination in relativity, and the orbital frequency is scaled in the same way). As before, $k_l$ is the static Love number and we have introduced
 two free parameters, $\delta$ and $\epsilon$. The first of these,  $\delta$, accounts for the gravitational redshift and the rotational frame-dragging induced by the orbital motion. However, it turns out that simply ``removing'' the redshift from the calculated mode frequency \cite{2016PhRvD..94j4028S} by taking $\delta = 1 - 2\mathcal C$ works quite well, so we are left with a  one-parameter expression for the dynamical tide. The example in figure~\ref{ltest} shows that the result
 compares well with a more thorough analysis in the effective-one-body framework \cite{2016PhRvL.116r1101H,2016PhRvD..94j4028S}. Moreover, the dynamical tide has been shown to compare well with results extracted from numerical simulations \cite{2019PhRvD..99d4008F}. This suggests that we have a good understanding of the physics of the dynamical tide.
 
 \begin{figure}
\includegraphics[width=0.7\columnwidth]{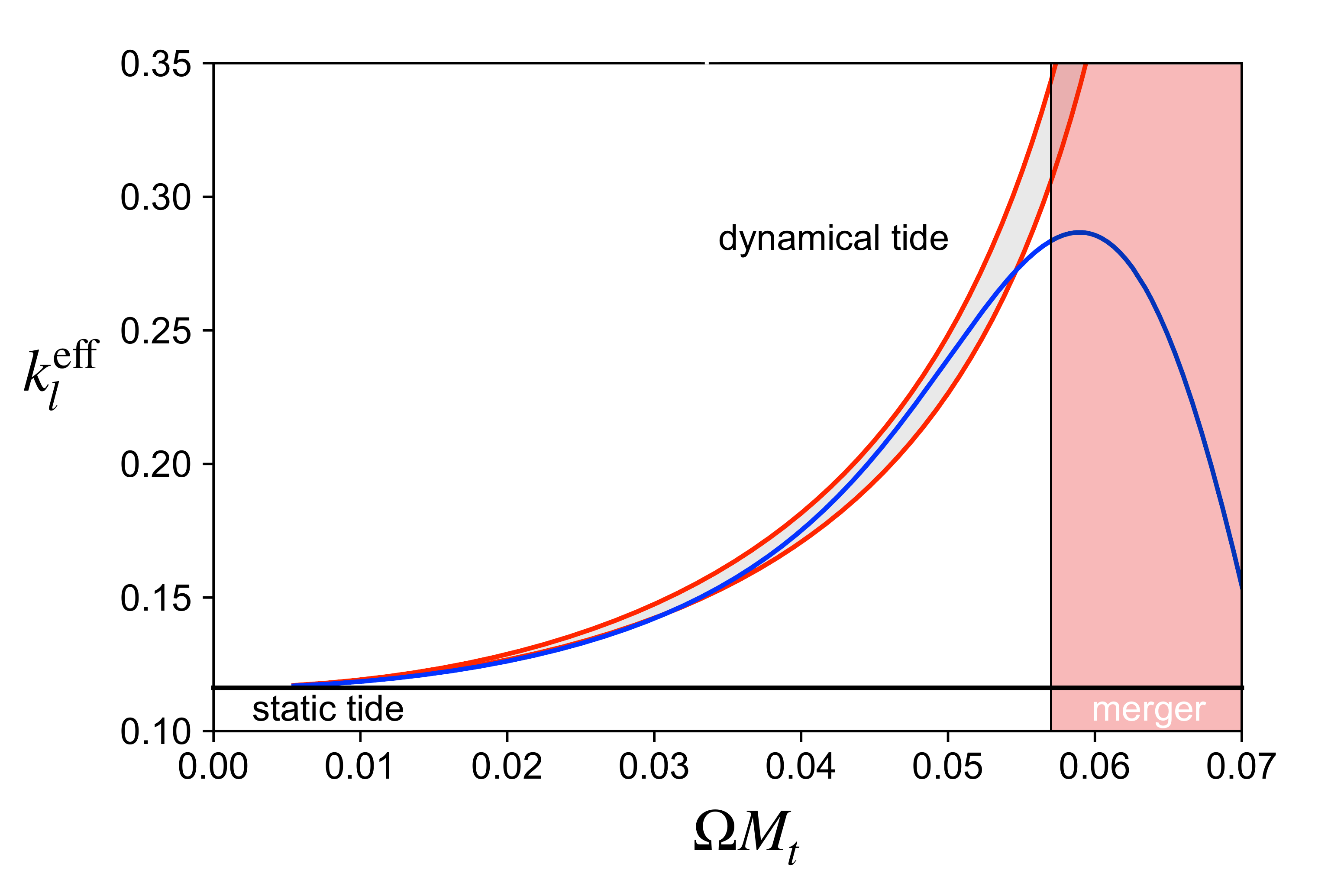}
\caption{Comparing the effective Love number $k_l^\mathrm{eff}$ from \eqref{kleff3}, using $\delta = 1- 2 \mathcal C$, to the results from \cite{2016PhRvD..94j4028S} for the quadrupole ($l=2$) contribution to the dynamical tide. The  horizontal line represents the static Love number ($k_l=k_l^\mathrm{eff}$ in the $\Omega\to 0 $ limit). The result from \cite{2016PhRvD..94j4028S} is shown as a solid blue curve, while estimates from \eqref{kleff3} are shown for the range $\epsilon=0.85-0.9$ with the latter representing the lower edge of the filled (grey) band. The region beyond the merger frequency, $\Omega M_t\approx 0.057$, with $M_t$ the total mass of the binary, is indicated by  the shaded region. } 
\label{ltest}
\end{figure}

Finally, we can  make contact with the end point of the binary evolution, the merger. 
The merger is associated with a wildly oscillating, hot remnant that (likely) eventually collapses to form a black hole.
Interestingly (given the present context), numerical simulations  demonstrate that the dynamics of the merged object exhibits robust high-frequency features \cite{2014PhRvD..90b3002B,2014PhRvL.113i1104T,2016CQGra..33h5003C}. These are generally thought to be associated with specific modes of oscillation, with the main feature associated with the fundamental mode \cite{2016EPJA...52...56B,2016CQGra..33h5003C,2015PhRvL.115i1101B}. This would be natural, although  not straightforward to demonstrate as we  have  to explore perturbations relative to a background that evolves on a relatively short timescale. 

From a seismology point of view, there is no reason to expect the merger dynamics to be connected with the inspiral phase. The physics of the equation of state should be ``the same'', but  the two inpiralling objects are cold (on the nuclear physics temperature scale) while the merged object is hot (possibly reaching temperatures higher than that of a supernova core) and differentially rotating. Since the thermal pressure and the rotation will both impact on the dynamics of the object \cite{2020MNRAS.497.5480C} one would not expect features of the two stages to be related in a simple way. Nevertheless, they seem to be. Focussing on an equal-mass binary and introducing \cite{2015PhRvL.115i1101B}
\begin{equation}
\kappa_2^t = \frac{1}{8} \frac{k_2}{C^5} = {3\over 16} \Lambda_2 \ ,
\end{equation}
 one finds (empirically) that the main  peak in the spectrum the merger dynamics ($f_2$) scales as 
\begin{equation}\label{f-mode_postmerg}
 f_2 M_t \approx 0.144  (\kappa_2^t)^{-0.278},
\end{equation}
where $M_t$ is the total mass of the system (as before; here assuming that we can neglect mass ejected during the merger). This relation appears to link the two stages of binary evolution, which could be useful when we (eventually) try to constrain the equation of state with observed merger signals. Hence, it makes sense to try to understand if and why the result should be robust.
We can make some progress towards an answer by noting that the  power law is similar to what one would expect for the  cold matter f-mode of the individual stars \cite{2020MNRAS.497.5480C}, which makes sense as it would associate the merger oscillation with the fundamental mode of the remnant, but the issue is not settled at this point. 

\section{Adding a bit of spin: The CFS instability}

We know from the several thousand observed radio pulsars that neutron stars tend to be spinning, in some cases with close to one revolution every millisecond. Due to the exquisite precision of radio timing the spin rate is---by a considerable margin---the most accurately determined neutron star parameter. Fundamental issues concerning the origin and evolution of the neutron star spin \cite{1998Natur.393..139S}, including irregularities like glitches and timing noise, continue to be explored. Rotation also impacts on neutron star seismology; it alters existing oscillation modes and introduces new ones \cite{2019gwa..book.....A}. In particular, the Coriolis force leads to the presence of so-called inertial modes \cite{1999ApJ...521..764L}. The r-mode---which may well be of  astrophysical relevance, but will not be discussed here (see chapter~15 in \cite{2019gwa..book.....A} for a detailed exposition)---is a particular example of this family of modes. In contrast, the centrifugal force does not introduce new families of modes, but as it deforms the shape of the star it can have significant impact on existing modes, shifting the frequencies as the spin rate increases.

Keeping our focus firmly on the f-mode, rotation breaks the symmetry of the non-rotating problem in such a way that the various $-l\le m\le l$ 
contributions become distinct and  rotation couples the different $l$-multipoles. In effect, each mode is affected differently by the star's spin.
As the rotation rate increases, an increasing number of  $Y_{l}^m$'s
are needed to  describe a given mode. One must also account for coupling
between the polar (with displacement vectors of form \eqref{displace}) and axial vectors (representing contributions proportional to $\epsilon^{ijk}(\nabla_j r) \nabla_k Y_l^m$, which are orthogonal to \eqref{displace}). These factors make the problem of calculating
pulsation modes of rapidly rotating stars a challenge. 
For rapidly spinning stars a frequency domain calculation based on a slow-rotation expansion becomes very messy when we go to second order (as many multipoles come into play). Relativity adds further complications as we have to account for the rotational frame-dragging, which effectively enters the problem as a differential rotation. The required outgoing-wave boundary condition that defines the relativistic quasinormal modes is also tricky\footnote{Because the rotating neutron star exterior is not associated with a Petrov type D spacetime \cite{2005MNRAS.358..923B} there is no (yet) known way to separate the variables, as in the case of the Teukolsky equation for spinning black holes.}. Because of these complications, much of the recent work has focussed on numerical simulations of the perturbation equations (see, for instance \cite{2002MNRAS.334..933J,2008PhRvD..78f4063G,2009MNRAS.394..730P,2009PhRvD..80f4026G,2011PhRvL.107j1102G}). Such time evolutions come with their own challenges. In particular, there is no easy way to isolate individual modes---in a sense you have to live with what you get from the chosen initial data and the (inevitable) numerical noise. Features like the f-mode may be relatively easy to identify, but higher order modes or aspects associated with the star's interior (like the crust or superfluid components) significantly less so.  To progress beyond the level of the current models we may need new ideas...

An important concept in the study of oscillating rotating stars is the 
pattern speed of a given mode. 
As each mode is proportional
to $e^{i(m\varphi+\omega t)}$, surfaces of constant phase are such that
\be
{d\varphi \over dt}= -{\omega \over m} = \sigma_p \ ,
\label{pattern}\ee
 defines the pattern speed, $\sigma_p$.
In terms of this quantity, we can make
 two observations for the ($l=m$) f-modes. Let us denote the 
mode frequency observed in the rotating frame by $\omega_r$, while the 
inertial frame frequency is $\omega_i$. We then
we see from  (\ref{fmode})  that the frequency of the f-modes 
increases with $m$ roughly as $\omega_r \sim \sqrt{m}$. 
According to (\ref{pattern}) this means
that  the pattern speed of the f-modes  
decreases as we increase $m$. As a consequence, one can always find an f-mode 
with arbitrarily  small pattern speed (corresponding to a  suitably 
large value of $m$) even though the high-order f-modes
have increasingly high frequencies. This will turn out to be an important observation.

It is also useful to note  that mode 
patterns corresponding to the different signs of $m$ tend to
rotate around the star in  opposite directions. Taking the
positive direction to be associated with the rotation of the star we 
see that the $l=\pm m$ modes are
backwards and forwards moving (retro/prograde), 
respectively, in the limit of vanishing rotation. 
However, rotation may change the situation. A very rough estimate of the corresponding 
mode for a rotating star (observed in the inertial frame) would be
\begin{equation}
\omega_i(\Omega) \approx  \omega_r(\Omega=0)   - m\Omega 
+ O(\Omega^2)  \ ,
\label{rotfreq}\end{equation} 
where $\Omega$ is the star's rotation rate. From 
 \eqref{fmode}  it then is easy to see that an originally retrograde mode may be dragged forwards to become prograde at some rotation rate. An observer on the star would still associate the oscillation with a wave moving ``backwards'', but an inertial observer would see the wave moving forwards, along with the bulk rotation. The importance of this will soon be clear. 

The two features we have sketched are important because the lay the foundation for the so-called CFS (Chandrasekhar-Friedman-Schutz) instability \cite{1970ApJ...161..561C,1978ApJ...222..281F}, a mechanism through which the emission of gravitational waves amplifies a given oscillation leading to enhanced emission and  a runaway process (at least until nonlinear effects become prominent \cite{2002PhRvD..65b4001S,2003ApJ...591.1129A,2015PhRvD..92h4018P}). It is important to understand how this works, so let us outline the argument. 

In the rotating case, where the background velocity $v^i$ does not vanish, the Lagrangian displacement is determined by and equation of form \cite{1978ApJ...221..937F,1978ApJ...222..281F}
\be
A \partial_t^2 \xi + B \partial_t \xi + C \xi = 0 \ .
\label{roteul}
\ee
where the symmetry properties for $A$ and $C$ remain as before, but
\be
B \partial_t \xi =
2\rho v^j \nabla_j \partial_t \xi_i \ , 
\ee
leads to
\be
\left<\eta,B\xi\right> = -\left<\xi,B\eta\right>^* \ .
\ee

Assuming that $\eta$ and $\xi$ both solve the perturbed Euler equation (\ref{roteul}),  again as before, it is relatively easy to show that the 
quantity
\be
W(\eta,\xi) = \left< \eta, A\partial_t \xi + { 1 \over 2}B\xi\right>
- \left< A\partial_t \eta + { 1 \over 2}B \eta, \xi\right> \ ,
\label{Wdef}\ee 
is conserved. That is, we have
\be
\partial_t W = 0  \ .
\ee
This 
motivates the definition of the  canonical energy 
\be
E_c = { 1 \over 2} W(\partial_t \xi,\xi) = { 1 \over 2} \left[
\left< \partial_t \xi , A \partial_t\xi \right>
+\left< \xi , C \xi \right> 
\right]\ ,
\label{CEdef}\ee
and, for axisymmetric systems (like a rotating star), the canonical angular momentum
\be
J_c = { 1\over 2} W(\partial_\varphi \xi, \xi) \ .
\ee
Both these quantities are conserved. 

The importance of the canonical energy and angular momentum stems from the fact that 
they can be used to test the stability of the system.
In particular \cite{1978ApJ...222..281F}:
\begin{itemize}
\item[(i)]  if the system is coupled to radiation (e.g. gravitational waves) 
which carries away positive energy  (which we take to mean that
$\partial_t E_c < 0$), then any initial data 
for which $E_c<0$ will lead to an instability. 

\item[(ii)] dynamical instabilities (not requiring additional physics like dissipation or wave emission) are only possible if $E_c=0$.
This is quite intuitive  
since the amplitude of 
a mode for which $E_c$ vanishes can grow 
without bound without violating any  conservation laws.

\end{itemize}

Let us now make contact with the seismology problem by considering a complex normal-mode
solution to the perturbation equations (which is convenient as it allows us to ``ignore'' the complicating presence of the so-called trivial displacements \cite{1978ApJ...222..281F}). That is, we assume a solution of form
\be
\xi^j = \tilde \xi^j  e^{i\omega t} \ ,
\ee
with $\omega$ possibly complex. 
Then the  canonical energy becomes
\be
E_c = \omega \left[ \mbox{ Re } \omega \left<  \tilde {\xi} , A \tilde {\xi}\right> - { i \over 2} 
\left< \tilde{\xi} , B \tilde{\xi}\right> \right] \ ,
 \label{Ec}
\ee
where the expression in the bracket is easily shown to be real valued.
For the canonical angular momentum we get
\be
J_c = -m \left[  \mbox{ Re } \omega \left<  \tilde{\xi} , A  \tilde{\xi} \right> - { i \over 2} 
\left< \tilde{\xi} , B \tilde{\xi}\right> \right] \ .
\label{Jc}\ee 
Combining the two relations we see that, for real frequency 
modes we have
\be
E_c = - {\omega \over m} J_c = \sigma_p J_c \ ,
\label{EJrel}\ee
where $\sigma_p$ is the pattern speed of the mode. This is an important relation as it connects the mode energy with the intuition associated with the pattern speed.

Moving on to the instability argument, for  real-frequency modes (and uniform rotation) we can use \eqref{Jc} to argue that \cite{1978ApJ...222..281F} 
\be
\sigma_p - \Omega \left( 1 + {1\over m} \right) \le { J_c/m^2 \over \left<\tilde\xi, \rho\tilde\xi \right>}
\le \sigma_p - \Omega \left( 1 - {1\over m} \right) \ .
\label{ineq1}\ee
Taking these inequalities as the starting point, consider a mode with 
finite frequency in the $\Omega \to 0$ limit (like the f-mode). For a co-rotating mode with $\sigma_p>0$ the relation \eqref{ineq1} implies that we must have $J_c>0$, while a counter-rotating mode
for which $\sigma_p < 0$ will have $J_c<0$. In both cases it follows from \eqref{EJrel} that $E_c>0$ , which means 
that both modes are stable. Now suppose  we have a mode such that $\sigma_p=0$ for a finite value of $
\Omega$. In a region near this point we must have
$J_c<0$. However, because of (\ref{EJrel}), 
$E_c$ will change sign at the point where $\sigma_p$ vanishes. As we have shown that the mode was stable
in the non-rotating limit the change of sign indicates that the canonical energy becomes negative, which represents the onset of instability. Basically,  a backwards moving mode is dragged forwards by
the rotation, as we have already discussed, but the mode is still moving 
backwards in the rotating frame. 
The gravitational waves from such a 
mode carry positive angular momentum 
away from the star but, since the perturbed fluid  
rotates slower than it would in  
absence of the perturbation, the angular momentum of the  mode is negative. The gravitational-wave emission 
makes the angular momentum 
increasingly negative and we have a runaway situation. The unstable mode continues to grow until viscosity or nonlinear effects come into play.

If we want to understand what this argument implies for realistic neutron star models, the logic is fairly straightforward (although the execution may be challenging). First we need to establish that the instability actually sets in at a realistic rotation rate. After all, the rotation of a fluid body is limited. Once the star rotates fast enough that it sheds mass at the equator, we cannot spin it up any more. This leads to the so called Kepler limit. Now, from what we have seen, we would expect the large $l=m$ f-modes to become unstable at a slower spin. However, these high order modes are associated with smaller scale fluid motion  so the viscous damping tends to dominate the gravitational-wave driving, thus ensuring that these modes remain stable \cite{1991ApJ...373..213I}. This complicates the problem---especially since we need to consider a range of possible dissipation mechanisms (like bulk and shear viscosity and the mutual friction that comes into play in a superfluid system \cite{1995ApJ...444..804L}). This is obviously difficult, as we do not really know the state and composition of the neutron star interior. Anyway, it is natural to start by trying to establish that the modes have the potential to be unstable in the first place, and worry about viscosity later. One way to do this is to focus on finding neutral modes \cite{1998ApJ...492..301S}, modes which have zero frequency in the inertial frame. Such stationary features  are easier to calculate and represent the system at the onset of the CFS instability. Detailed work on this problem \cite{1998ApJ...492..301S,1997ApJ...488..799K} shows that the quadrupole f-mode---which is unlikely to become unstable in a rotating Newtonian star---reaches the instability threshold just below the Kepler limit for a realistic neutron star model. This then opens a window of opportunity for the instability to play a role in astrophysical scenarios\footnote{From the astrophysics point of view, the CFS instability is relevant as it offers an explanation for the absence of observed systems spinning close to the Kepler limit \cite{2019gwa..book.....A}.} involving the most rapidly spinning stars. 

\begin{figure}
\includegraphics[width=0.7\columnwidth]{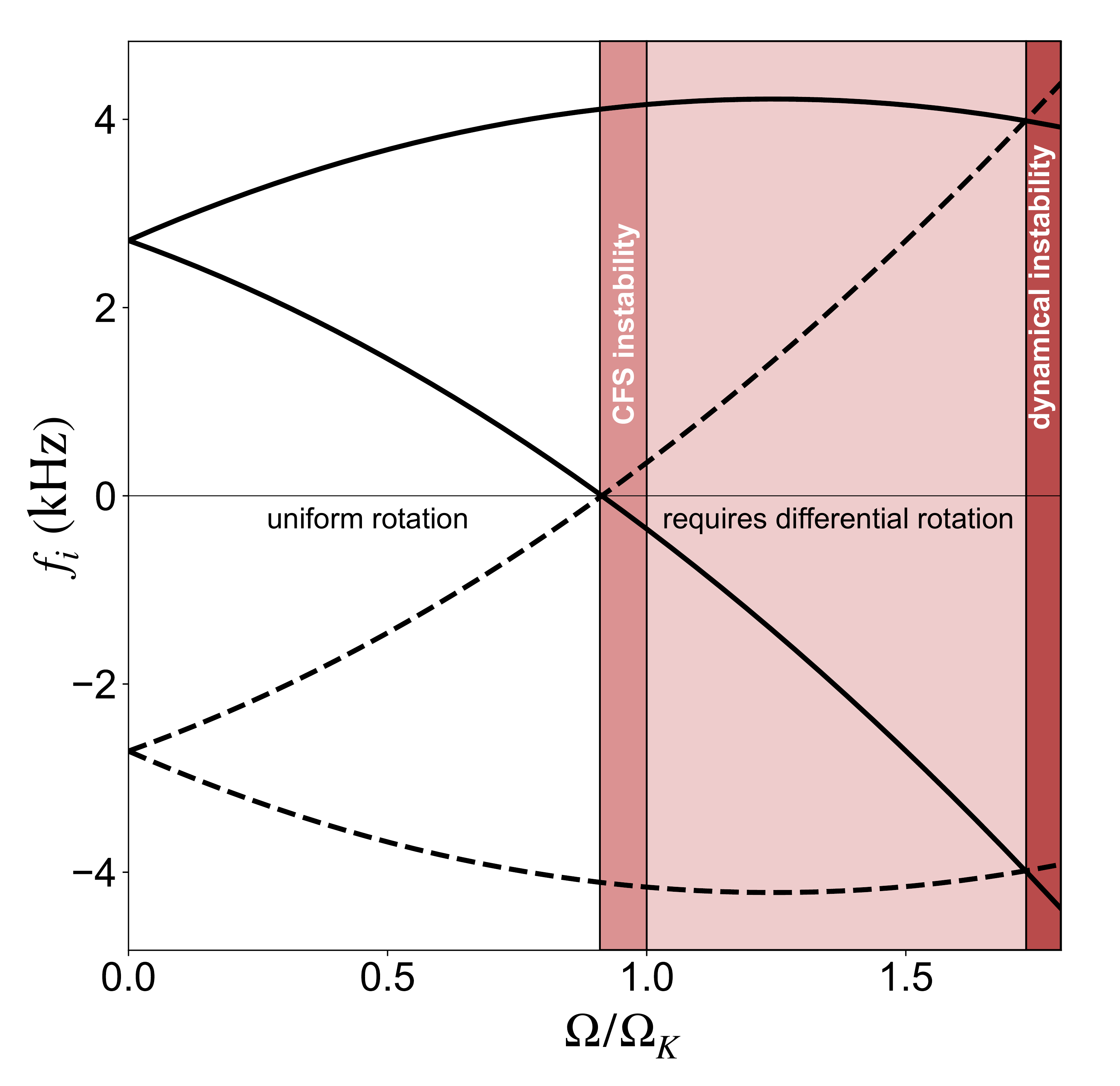} 
\caption{Illustrating the numerical results from \cite{2013PhRvD..88d4052D}  for the $l=2$ f-modes of rotating neutron stars. The inertial frame mode frequency $f_i=\omega_i/2\pi$ (in kHz) is shown as function of the (uniform) rotation frequency $\Omega$ (scaled to the Kepler frequency $\Omega_K$).  }
\label{ffig}
\end{figure}

If we want to consider the role of viscosity or consider the evolution of an unstable mode, we need dynamical mode calculation. The most detailed work in this direction builds on time-evolutions for rotating relativistic stars (typically in the Cowling approximation \cite{2008PhRvD..78f4063G}, where variations in the metric are ignored, although see \cite{2020PhRvL.125k1106K}).
As an illustration, let us consider the results from \cite{2013PhRvD..88d4052D} which provide empirical relations based on results for a collection of matter equations of state. For the quadrupole f-modes we have, first of all, the frequency of the mode for non-rotating stars\footnote{Note that this leads to the f-mode frequency $f\approx 2.7$~kHz, which is  consistent with the results we quoted earlier.}:
\begin{equation}
{1\over 2\pi} \omega_0\ \approx 1.562 + 1.151  \left( {M_0 \over 1.4M_\odot} \right)^{1/2} \left( {R_0\over 10\ \mathrm{km}}\right)^{-3/2} \   (\mathrm{kHz})\ , 
\end{equation}
where $M_0$ and $R_0$ are the mass and radius of the non-rotating model, respectively. With our conventions, the inertial frame frequency of the prograde  ($l=-m=2$ in the non-rotating limit) mode is (combining the results from \cite{2013PhRvD..88d4052D} with  $\omega_i = \omega_r - m\Omega$):
\begin{equation}\label{f-mode_inertial_stable}
\omega_i = \omega_0 \left[ 1 - 0.235 \left( {\Omega \over \Omega_K}\right)-\frac{m\Omega}{\omega_0} -0.358 \left( {\Omega\over \Omega_K}\right)^2 \right] \ .
\end{equation}
In this expression, the Kepler frequency, $\Omega_K$, is estimated by 
\begin{equation}
{1\over 2\pi} \Omega_K\ \approx 1.716 \left( {M_0 \over 1.4M_\odot} \right)^{1/2} \left( {R_0\over 10\ \mathrm{km}}\right)^{-3/2} - 0.189 \  (\mathrm{kHz}) \ .
\end{equation}
 Similarly, for the retrograde ($l=m=2$) f-mode we have
\begin{equation}\label{f-mode_inertial_unstable}
\omega_i = \omega_0 \left[ 1 +0.402 \left( {\Omega \over \Omega_K}\right)-\frac{m\Omega}{\omega_0} -0.406 \left( {\Omega\over \Omega_K}\right)^2 \right] \ .
\end{equation}

In order to understand the implications of these results, we may consider the illustration in figure~\ref{ffig}, which shows the behaviour of the modes that limit to $\pm\omega_0$ in the slow-rotation limit (the rotational splitting leads to a pair of modes for each  $m=\pm l$).  Focussing, first of all, on the pair of modes that start out with a positive frequency, we see that the frequency of the prograde $l=-m=2$ mode increases with the rotation rate. Meanwhile, the observed inertial frame frequency of the originally retrograde mode decreases and passes through zero just above $\Omega \approx 0.9 \Omega_K$. At this point the pattern speed of the mode changes sign, signalling the onset of the CFS instability. As expected, the star has to spin very fast for the instability to kick in. Extrapolating beyond the region of validity of the results, which is cheating\footnote{Models reaching beyond  $\Omega_K$ require differential rotation.} but serves the purpose of illustrating another important point, we see that the CFS unstable mode that started out with negative frequency eventually catches up with the originally prograde mode. The point where these two modes merge indicate the onset of dynamical instability---the so-called bar-mode instability \cite{2000PhRvD..62f4019N,2000ApJ...542..453S,2007PhRvD..75d4023B}. This dynamical instability requires $E_c=0$ which is possible when the two modes merge. This behaviour (mode merger leading to instability) is familiar from many wave problems. Beyond this point our extrapolation absolutely does not make sense\footnote{Up to the mode merger the extrapolation can be ``justified'' by results for rotating ellipsoids, see \cite{1973efe..book.....C} and the discussion in \cite{2019gwa..book.....A}.}, as the pair of modes would be replaced by a single complex-valued frequency once the dynamical instability sets in.  

While we have no observational evidence for mature neutron stars spinning fast enough for the f-mode instability to come into play, it may be relevant for (some fraction of) newly born systems. This depends entirely on the birth-spin expected from a supernova core collapse. This issue is not particularly well understood but it is expected \cite{1998Natur.393..139S} that some fraction of neutron stars will be born rapidly spinning. This may, in fact, be a requirement for the dynamo mechanism needed to wind up the star's magnetic field to magnetar strength \cite{1995MNRAS.275..255T} (but, again, the detailed mechanism for this is not understood). 
The unstable f-mode may also impact on the evolution of massive post-merger remnants \cite{2015PhRvD..92j4040D}. The outcome depends on, first of all, whether the spin of the system will be sufficient for the instability to be triggered and, secondly, the saturation amplitude of the unstable mode (expected to be due to the nonlinear coupling to other modes \cite{2002PhRvD..65b4001S,2003ApJ...591.1129A,2015PhRvD..92h4018P}). It has been suggested  that this kind of  signal may  be detectable at the distance to the Virgo cluster  \cite{2015PhRvD..92j4040D}, an idea worth further consideration.

\section{Final comments}

In this brief overview, we have discussed  aspects of neutron star seismology and how the problem connects with the rapidly developing area of gravitational-wave astronomy. The aim was to introduce the ideas rather than provide an exhaustive discussion of every possible angle. Hence, the  fundamental f-mode was allowed to take centre stage throughout, at the risk of downplaying the role of other modes---like the infamous r-mode---and perhaps giving an overly simplified picture of the  oscillation problem. Given this, it makes sense to conclude with some comments on the bigger picture---what we know and what we don't know.

First of all, we are painfully aware that the oscillation problem for realistic neutron star models is  complicated. Different aspects of physics tend to be associated with (more or less) distinct families of oscillation modes---composition/entropy gradients support the g-modes, the Coriolis force leads to inertial modes, there are elastic shear modes in the crust, there are superfluid modes and so on. The range of physics can be intimidating and the progress on modelling these systems in as much detail as we can muster has been fairly slow.
Nevertheless, there have been notable advances.

Before going into the details it is worth going over a couple of overarching ``ground rules''. 
First of all, it is important to keep track of the ultimate aim of the exercise: We want to combine our theory models with observations to constrain  the physics. This means that---for each conceived scenario and observation channel--- we must not lose track of what may be achievable with current and foreseeable technology. From the perpective of gravitational-wave astronomy (and given that neutron star dynamics tends to be a high-frequency (kHz) phenomenon) our speculation should focus on the current ground-based instruments and what may be possible with  different planned  3G detectors like the Einstein Telescope or a Cosmic Explorer. In fact, clear arguments that certain aspects of neutron star physics will come within reach in the 3G era would add to the---already  strong \cite{2020JCAP...03..050M}---science case for these instruments.  

Secondly, we need to develop the models in the framework of general relativity (or whatever fashionable alternative theory may take our fancy). The motivation for this is simple: A realistic matter description only makes sense in the context of a relativistic stellar model. For example, it is well known that, for a given central density, the radius of a Newtonian star differs considerably from the corresponding relativistic model. The ``errors'' tend to be so large that any discussion of the impact of the fine print physics becomes meaningless. This does not mean that we have not learned valuable lessons from Newtonian models. On the contrary, much of our understanding of the fluid dynamics involved draws on work in the non-relativistic regime---and will likely continue to do so for some time. 

With these two points in mind, let us take stock of where we stand. When it comes to studying the oscillation properties of ``realistic'' neutron star models, the state of the art models in \cite{2015PhRvD..92f3009K}  combine internal composition gradients with realistic cooling (including the freezing of the star's crust) to explore the evolution of the oscillation spectrum as a young neutron star matures. These models do not account for superfluity or, indeed, the magnetic field. The former aspect is not an insurmountable challenge, given the formalism discussed in \cite{2019CQGra..36j5004A} and work on the torsional oscillations of the crust (motivated by the observed quasiperiodic features in the tail of some magnetar flares \cite{2009CQGra..26o5016S} but the magnetic field provides a technical hurdle. This is not because we are unable to study the oscillations of magnetic stars---the results from \cite{2012MNRAS.421.2054G,2012MNRAS.423..811C} show that such work is well under way---but the simple reality is that the internal magnetic field configuration (which we need to provide a background for our perturbation analysis) is not well understood. In fact, this is a drastic understatement. None of the models we are able to build appear to be stable in the long term \cite{2012MNRAS.424..482L,2020MNRAS.499.2636B} and this presents us with an obvious conceptual problem. In short, we are still some way away from modelling (even non-spinning) neutron stars in their full glory. It is important that strive to do so, because the different physics aspects are not independent and we need to understand how they are linked if we want make secure inference statements for observed signals. This is particularly the case for the dynamical tide which has so far   been considered only for Newtonian models. This is an issue that need attention.

When it comes to rotating relativistic stars, our models are much less complete. We have a fairly good handle on the impact of rotation on the f-mode and some idea of the effect on the pressure p-modes and the gravity g-modes \cite{2009PhRvD..80f4026G}. This work provides an understanding of the onset of the CFS instability associated with the f-mode (for very high spin rates) \cite{2008PhRvD..78f4063G,2010PhRvD..81h4019K,2011PhRvL.107j1102G,2013PhRvD..88d4052D,2020PhRvL.125k1106K}, but we do not yet have a good handle on the many different dissipation mechanism that may affect an unstable mode. There is a sharp contrast between the physics that has been brought into play in efforts to ``kill'' the inertial r-mode instability (see \cite{2019gwa..book.....A}) and the level of discussion for the f-mode. Having said that, the work on the r-modes is almost exclusively in the Newtonian domain so we are lacking in that respect, as well (see \cite{2001PhRvD..63b4019L,2003PhRvD..68l4010L,2015PhRvD..91b4001I} for a few notable exceptions). Similar statements apply to the effort to understand the saturation of unstable modes, through the nonlinear coupling to other modes \cite{2002PhRvD..65b4001S,2003ApJ...591.1129A,2015PhRvD..92h4018P}, for which a  relativistic framework has not been developed. And let us not forget the phenomenology that comes into play when we add differential rotation to the mix.  In particular, suggestions that the so-called low-T/W instability \cite{2001ApJ...550L.193C,2005ApJ...618L..37W} may act in merger remnants are intriguing \cite{2020MNRAS.498.5904P,2020PhRvD.102d4040X,2020PhRvD.101f4052D}. Again, this is a mechanism that is much less well understood than it ought to be. The ``wish list'' for rotating stars becomes quite long...

A much briefer summary would simply state that we  have a lot of work left to do.

\acknowledgments{My effort to understand neutron star seismology has involved fruitful discussions with many great collaborators. I would  single out Bernard Schutz, who pointed me in this direction in the first place, and Kostas Kokkotas, who taught me what was what. Without the two of them, I may well have spent the last 30 years or so thinking about something else. }



\begin{thebibliography}{999}

\bibitem[Demorest \em{et~al.}(2010)Demorest et~al.]{2010Natur.467.1081D}
Demorest, P.; et al
\newblock {A two-solar-mass neutron star measured using Shapiro delay}.
\newblock { Nature} {\bf 2010}, {\em 467},~1081.

\bibitem[{Riley} \em{et~al.}(2019){Riley}, {Watts}, {Bogdanov}, {Ray},
  {Ludlam}, {Guillot}, {Arzoumanian}, {Baker}, {Bilous}, {Chakrabarty},
  {Gendreau}, {Harding}, {Ho}, {Lattimer}, {Morsink}, and
  {Strohmayer}]{2019ApJ...887L..21R}
{Riley}, T.E.; et al
\newblock {A NICER View of PSR J0030+0451: Millisecond Pulsar Parameter
  Estimation}.
\newblock { Ap. J. Lett.} {\bf 2019}, {\em 887},~L21.
 

\bibitem[{Miller} \em{et~al.}(2019){Miller}, {Lamb}, {Dittmann}, {Bogdanov},
  {Arzoumanian}, {Gendreau}, {Guillot}, {Harding}, {Ho}, {Lattimer}, {Ludlam},
  {Mahmoodifar}, {Morsink}, {Ray}, {Strohmayer}, {Wood}, {Enoto}, {Foster},
  {Okajima}, {Prigozhin}, and {Soong}]{2019ApJ...887L..24M}
{Miller}, M.C.; et al
\newblock {PSR J0030+0451 Mass and Radius from NICER Data and Implications for
  the Properties of Neutron Star Matter}.
\newblock {Ap. J. Lett.} {\bf 2019}, {\em 887},~L24.

\bibitem[{Watts} \em{et~al.}(2016){Watts} et~al.]{2016RvMP...88b1001W}
{Watts}, A.L.; et al
\newblock {Colloquium: Measuring the neutron star equation of state using X-ray
  timing}.
\newblock { Reviews of Modern Physics} {\bf 2016}, {\em 88},~021001.

\bibitem[{Andersson}(2019)]{2019gwa..book.....A}
{Andersson}, N.
\newblock {\em {Gravitational-Wave Astronomy: Exploring the Dark Side of the
  Universe}}; Oxford University Press, Oxford, 2019.

\bibitem[{Aerts} \em{et~al.}(2010){Aerts}, {Christensen-Dalsgaard}, and
  {Kurtz}]{2010aste.book.....A}
{Aerts}, C.; {Christensen-Dalsgaard}, J.; {Kurtz}, D.W.
\newblock {\em {Asteroseismology}}; Springer, Heidelberg,  2010.

\bibitem[{Andersson} and {Kokkotas}(1996)]{1996PhRvL..77.4134A}
{Andersson}, N.; {Kokkotas}, K.D.
\newblock {Gravitational Waves and Pulsating Stars: What Can We Learn from
  Future Observations?}
\newblock {Phys. Rev. Lett.} {\bf 1996}, {\em 77},~4134.

\bibitem[{Andersson} and {Kokkotas}(1998)]{1998MNRAS.299.1059A}
{Andersson}, N.; {Kokkotas}, K.D.
\newblock {Towards gravitational-wave asteroseismology}.
\newblock {MNRAS} {\bf 1998}, {\em 299},~1059.

\bibitem[{Kokkotas} \em{et~al.}(2001){Kokkotas}, {Apostolatos}, and
  {Andersson}]{2001MNRAS.320..307K}
{Kokkotas}, K.D.; {Apostolatos}, T.A.; {Andersson}, N.
\newblock {The inverse problem for pulsating neutron stars: a `fingerprint
  analysis' for the supranuclear equation of state}.
\newblock {MNRAS} {\bf 2001}, {\em 320},~307.

\bibitem[{Benhar} \em{et~al.}(2004){Benhar}, {Ferrari}, and
  {Gualtieri}]{2004PhRvD..70l4015B}
{Benhar}, O.; {Ferrari}, V.; {Gualtieri}, L.
\newblock {Gravitational-wave asteroseismology reexamined}.
\newblock {Phys. Rev. D} {\bf 2004}, {\em 70},~124015.

\bibitem[{Friedman} and {Schutz}(1978)]{1978ApJ...221..937F}
{Friedman}, J.L.; {Schutz}, B.F.
\newblock {Lagrangian perturbation theory of nonrelativistic fluids}.
\newblock {Ap. J.} {\bf 1978}, {\em 221},~937.

\bibitem[{Detweiler}(1975)]{1975ApJ...197..203D}
{Detweiler}, S.L.
\newblock {A variational calculation of the fundamental frequencies of
  quadrupole pulsation of fluid spheres in general relativity}.
\newblock {Ap. J.} {\bf 1975}, {\em 197},~203.

\bibitem[{Andersson} \em{et~al.}(1995){Andersson}, {Kokkotas}, and
  {Schutz}]{1995MNRAS.274.1039A}
{Andersson}, N.; {Kokkotas}, K.D.; {Schutz}, B.F.
\newblock {A new numerical approach to the oscillation modes of relativistic
  stars}.
\newblock {MNRAS} {\bf 1995}, {\em 274},~1039.

\bibitem[{Leins} \em{et~al.}(1993){Leins}, {Nollert}, and
  {Soffel}]{1993PhRvD..48.3467L}
{Leins}, M.; {Nollert}, H.P.; {Soffel}, M.H.
\newblock {Nonradial oscillations of neutron stars: A new branch of strongly
  damped normal modes}.
\newblock {Phys. Rev. D} {\bf 1993}, {\em 48},~3467.

\bibitem[{Glampedakis} and {Andersson}(2003)]{2003CQGra..20.3441G}
{Glampedakis}, K.; {Andersson}, N.
\newblock {'Quick and dirty' methods for studying black-hole resonances}.
\newblock {Class. Quantum Grav.} {\bf 2003}, {\em 20},~3441.

\bibitem[{Lattimer} and {Prakash}(2001)]{2001ApJ...550..426L}
{Lattimer}, J.M.; {Prakash}, M.
\newblock {Neutron Star Structure and the Equation of State}.
\newblock {Ap. J.} {\bf 2001}, {\em 550},~426.

\bibitem[{Tolman}(1939)]{1939PhRv...55..364T}
{Tolman}, R.C.
\newblock {Static solutions of Einstein's field equations for spheres of
  fluid}.
\newblock { Physical Review} {\bf 1939}, {\em 55},~364.

\bibitem[{Yagi} and {Yunes}(2013)]{2013Sci...341..365Y}
{Yagi}, K.; {Yunes}, N.
\newblock {I--Love--Q: Unexpected universal relations for neutron stars and
  quark stars}.
\newblock { Science} {\bf 2013}, {\em 341},~365.

\bibitem[{Tsui} and {Leung}(2005)]{2005MNRAS.357.1029T}
{Tsui}, L.K.; {Leung}, P.T.
\newblock {Universality in quasi-normal modes of neutron stars}.
\newblock {MNRAS} {\bf 2005}, {\em 357},~1029.

\bibitem[{Lau} \em{et~al.}(2010){Lau}, {Leung}, and {Lin}]{2010ApJ...714.1234L}
{Lau}, H.K.; {Leung}, P.T.; {Lin}, L.M.
\newblock {Inferring physical parameters of compact stars from their f-mode
  gravitational-wave signals}.
\newblock {Ap. J.} {\bf 2010}, {\em 714},~1234.

\bibitem[{Friedman} and {Schutz}(1978)]{1978ApJ...222..281F}
{Friedman}, J.L.; {Schutz}, B.F.
\newblock {Secular instability of rotating Newtonian stars}.
\newblock {Ap. J.} {\bf 1978}, {\em 222},~281.

\bibitem[{Gundlach} \em{et~al.}(1994){Gundlach}, {Price}, and
  {Pullin}]{1994PhRvD..49..883G}
{Gundlach}, C.; {Price}, R.H.; {Pullin}, J.
\newblock {Late-time behavior of stellar collapse and explosions. I. Linearized
  perturbations}.
\newblock {Phys. Rev. D} {\bf 1994}, {\em 49},~883.

\bibitem[{Burrows} \em{et~al.}(2006){Burrows}, {Livne}, {Dessart}, {Ott}, and
  {Murphy}]{2006NewAR..50..487B}
{Burrows}, A.; {Livne}, E.; {Dessart}, L.; {Ott}, C.D.; {Murphy}, J.
\newblock {An acoustic mechanism for core-collapse supernova explosions}.
\newblock { New Astronomy Reviews} {\bf 2006}, {\em 50},~487.

\bibitem[{Radice} \em{et~al.}(){Radice}, {Morozova}, {Burrows}, {Vartanyan},
  and {Nagakura}]{2019ApJ...876L...9R}
{Radice}, D.; {Morozova}, V.; {Burrows}, A.; {Vartanyan}, D.; {Nagakura}, H.
\newblock {Characterizing the Gravitational Wave Signal from Core-collapse
  Supernovae}.
\newblock {Ap. J. Lett.} {\bf 2019}, {\em 876},~L9.

\bibitem[{Blaes} \em{et~al.}(1989){Blaes}, {Blandford}, {Goldreich}, and
  {Madau}]{1989ApJ...343..839B}
{Blaes}, O.; {Blandford}, R.; {Goldreich}, P.; {Madau}, P.
\newblock {Neutron Starquake Models for Gamma-Ray Bursts}.
\newblock {Ap. J.} {\bf 1989}, {\em 343},~839.


\bibitem[{Mock} and {Joss}(1998)]{1998ApJ...500..374M}
{Mock}, P.C.; {Joss}, P.C.
\newblock {Limits on Energy Storage in the Crusts of Accreting Neutron Stars}.
\newblock {Ap. J.} {\bf 1998}, {\em 500},~374.

\bibitem[{Barat} \em{et~al.}(1983){Barat} et~al.]{1983A&A...126..400B}
{Barat}, C.; et al
\newblock {Fine time structure in the 1979 March 5 gamma-ray burst}.
\newblock { Astron. Astrophys.} {\bf 1983}, {\em 126},~400.

\bibitem[{Samuelsson} and {Andersson}(2007)]{2007MNRAS.374..256S}
{Samuelsson}, L.; {Andersson}, N.
\newblock {Neutron star asteroseismology. Axial crust oscillations in the
  Cowling approximation}.
\newblock { MNRAS} {\bf 2007}, {\em 374},~256.

\bibitem[{Baggio} \em{et~al.}(2005){Baggio} et~al.]{2005PhRvL..95h1103B}
{Baggio}, L.; others.
\newblock {Upper limits on gravitational-wave emission in association with the
  27 Dec 2004 giant flare of SGR1806$-$20}.
\newblock { Phys. Rev. Lett.} {\bf 2005}, {\em 95},~081103.

\bibitem[Abbott \em{et~al.}(2007)Abbott et~al.]{2007PhRvD..76f2003A}
Abbott, B.P.; et al
\newblock {Search for gravitational wave radiation associated with the
  pulsating tail of the SGR 1806$-$20 hyperflare of 27 December 2004 using
  LIGO}.
\newblock { Phys. Rev. D} {\bf 2007}, {\em 76},~062003.

\bibitem[Abbott \em{et~al.}(2008)Abbott et~al.]{2008PhRvL.101u1102A}
Abbott, B.P.; et al.
\newblock {Search for gravitational-wave bursts from soft gamma repeaters}.
\newblock { Phys. Rev. Lett.} {\bf 2008}, {\em 101},~211102.

\bibitem[Abbott \em{et~al.}(2009)Abbott et~al.]{2009ApJ...701L..68A}
Abbott, B.P.; et al.
\newblock {Stacked search for gravitational waves from the 2006 SGR 1900$+$14
  storm}.
\newblock { Ap. J. Lett.} {\bf 2009}, {\em 701},~L68.

\bibitem[{Abadie} \em{et~al.}(2011){Abadie} et~al.]{2011ApJ...734L..35A}
{Abadie}, J.; et al.
\newblock {Search for gravitational wave bursts from six magnetars}.
\newblock { Ap. J. Lett.} {\bf 2011}, {\em 734},~L35.

\bibitem[{Haskell} and {Melatos}(2015)]{2015IJMPD..2430008H}
{Haskell}, B.; {Melatos}, A.
\newblock {Models of pulsar glitches}.
\newblock { International Journal of Modern Physics D} {\bf 2015}, {\em
  24},~1530008.

\bibitem[{Andersson} and {Comer}(2001)]{2001PhRvL..87x1101A}
{Andersson}, N.; {Comer}, G.L.
\newblock {Probing neutron-star superfluidity with gravitational-wave data}.
\newblock {Phys. Rev. Lett.} {\bf 2001}, {\em 87},~241101.

\bibitem[{Sidery} \em{et~al.}(2010){Sidery}, {Passamonti}, and
  {Andersson}]{2010MNRAS.405.1061S}
{Sidery}, T.; {Passamonti}, A.; {Andersson}, N.
\newblock {The dynamics of pulsar glitches: Contrasting phenomenology with
  numerical evolutions}.
\newblock {MNRAS} {\bf 2010}, {\em 405},~1061.

\bibitem{2011PhRvD..83d2001A}
{Abadie}, J.; et al.
\newblock {Search for gravitational waves associated with the August 2006
  timing glitch of the Vela pulsar}.
\newblock { Phys. Rev. D} {\bf 2011}, {\em 83},~042001.

\bibitem[{Chatterjee}(2019{\natexlab{a}})]{chatterjee19}
{Chatterjee}, D.
\newblock { GRB Coordinates Network} {\bf 2019}, {\em 26222},~1.

\bibitem[{Chatterjee}(2019{\natexlab{b}})]{chatterjee19b}
{Chatterjee}, D.
\newblock { GRB Coordinates Network} {\bf 2019}, {\em 26250},~1.

\bibitem[{Kaplan} and {Read}(2019)]{kaplan}
{Kaplan}, D., F.J.; {Read}, J.
\newblock { GRB Coordinates Network} {\bf 2019}, {\em 26243},~1.

\bibitem[{Ho} \em{et~al.}(2020){Ho}, {Jones}, {Andersson}, and
  {Espinoza}]{2020PhRvD.101j3009H}
{Ho}, W.C.G.; {Jones}, D.I.; {Andersson}, N.; {Espinoza}, C.M.
\newblock {Gravitational waves from transient neutron star f -mode
  oscillations}.
\newblock {Phys. Rev. D} {\bf 2020}, {\em 101},~103009.

\bibitem[{Reisenegger} and {Goldreich}(1992)]{1992ApJ...395..240R}
{Reisenegger}, A.; {Goldreich}, P.
\newblock {A new class of g-modes in neutron stars}.
\newblock { Ap. J.} {\bf 1992}, {\em 395},~240.

\bibitem[{Andersson} and {Pnigouras}()]{2019MNRAS.489.4043A}
{Andersson}, N.; {Pnigouras}, P.
\newblock {The g-mode spectrum of reactive neutron star cores}.
\newblock {MNRAS} {\bf 2019} {\em 489}, 4043.

\bibitem[{Haensel}(1992)]{1992A&A...262..131H}
{Haensel}, P.
\newblock {Non-equilibrium neutrino emissivities and opacities of neutron star
  matter}.
\newblock {Astron. Astrophys.} {\bf 1992}, {\em 262},~131.

\bibitem[{Reisenegger}(1995)]{1995ApJ...442..749R}
{Reisenegger}, A.
\newblock {Deviations from Chemical Equilibrium Due to Spin-down as an Internal
  Heat Source in Neutron Stars}.
\newblock {Ap. J.} {\bf 1995}, {\em 442},~749.

\bibitem[{Chamel} \em{et~al.}(2011){Chamel} et~al.]{2011PhRvC..84f2802C}
{Chamel}, N.; et al.
\newblock {Masses of neutron stars and nuclei}.
\newblock {Phys. Rev. C} {\bf 2011}, {\em 84},~062802.

\bibitem[{Schmitt} and {Shternin}(2018)]{2018ASSL..457..455S}
{Schmitt}, A.; {Shternin}, P., {Reaction Rates and Transport in Neutron Stars}.
\newblock In {\em Astrophysics and Space Science Library}; {Rezzolla}, L.;
  {Pizzochero}, P.; {Jones}, D.I.; {Rea}, N.; {Vida{\~n}a}, I., Eds.;  2018;
  Vol. 457, p. 455.

\bibitem[{Ferrari} \em{et~al.}(2003){Ferrari}, {Miniutti}, and
  {Pons}]{2003MNRAS.342..629F}
{Ferrari}, V.; {Miniutti}, G.; {Pons}, J.
\newblock {Gravitational waves from newly born, hot neutron stars}.
\newblock {MNRAS} {\bf 2003}, {\em 342},~629.

\bibitem[{Gualtieri} \em{et~al.}(2004){Gualtieri}, {Pons}, and
  {Miniutti}]{2004PhRvD..70h4009G}
{Gualtieri}, L.; {Pons}, J.A.; {Miniutti}, G.
\newblock {Nonadiabatic oscillations of compact stars in general relativity}.
\newblock {Phys. Rev. D} {\bf 2004}, {\em 70},~084009.

\bibitem[{Andersson}(2021)]{2021Univ....7...17A}
{Andersson}, N.
\newblock {A Superfluid Perspective on Neutron Star Dynamics}.
\newblock { Universe} {\bf 2021}, {\em 7},~17.

\bibitem[{Lee}(1995)]{1995A&A...303..515L}
{Lee}, U.
\newblock {Nonradial oscillations of neutron stars with the superfluid core}.
\newblock {Astron. Astrophys.} {\bf 1995}, {\em 303},~515.

\bibitem[{Andersson} and {Comer}(2001)]{2001MNRAS.328.1129A}
{Andersson}, N.; {Comer}, G.L.
\newblock {On the dynamics of superfluid neutron star cores}.
\newblock {MNRAS} {\bf 2001}, {\em 328},~1129.

\bibitem[{Gusakov} and {Kantor}(2013)]{2013PhRvD..88j1302G}
{Gusakov}, M.E.; {Kantor}, E.M.
\newblock {Thermal g-modes and unexpected convection in superfluid neutron
  stars}.
\newblock {Phys. Rev. D} {\bf 2013}, {\em 88},~101302.

\bibitem[{Passamonti} \em{et~al.}(2016){Passamonti}, {Andersson}, and
  {Ho}]{2016MNRAS.455.1489P}
{Passamonti}, A.; {Andersson}, N.; {Ho}, W.C.G.
\newblock {Buoyancy and g-modes in young superfluid neutron stars}.
\newblock {MNRAS} {\bf 2016}, {\em 455},~1489.

\bibitem[{Blanchet}(2006)]{2006LRR.....9....4B}
{Blanchet}, L.
\newblock {Gravitational radiation from post-Newtonian sources and inspiralling
  compact binaries}.
\newblock { Living Reviews in Relativity} {\bf 2006}, {\em 9},~4.

\bibitem[{Flanagan} and {Hinderer}(2008)]{2008PhRvD..77b1502F}
{Flanagan}, {\'E}.{\'E}.; {Hinderer}, T.
\newblock {Constraining neutron-star tidal Love numbers with gravitational-wave
  detectors}.
\newblock {Phys. Rev. D} {\bf 2008}, {\em 77},~021502.

\bibitem[{Hinderer}(2008)]{2008ApJ...677.1216H}
{Hinderer}, T.
\newblock {Tidal Love numbers of neutron stars}.
\newblock { Ap. J.} {\bf 2008}, {\em 677},~1216.

\bibitem[{Hinderer} \em{et~al.}(2010){Hinderer} et~al.]{2010PhRvD..81l3016H}
{Hinderer}, T.; et al.
\newblock {Tidal deformability of neutron stars with realistic equations of
  state and their gravitational-wave signatures in binary inspiral}.
\newblock { Phys. Rev. D} {\bf 2010}, {\em 81},~123016.

\bibitem[Abbott \em{et~al.}(2017)Abbott et~al.]{2017ApJ...848L..12A}
Abbott, B.P.; et al.
\newblock {Multi-messenger observations of a binary neutron-star merger}.
\newblock { Ap. J. Lett.} {\bf 2017}, {\em 848},~L12.

\bibitem[Abbott \em{et~al.}(2018)Abbott et~al.]{2018arXiv180511581T}
Abbott, B.P.; et al.
\newblock {GW170817: Measurements of neutron star radii and equation of state}.
\newblock {Phys. Rev. Lett.} {\bf 2018}, {\em 121},~161101.

\bibitem{2018PhRvL.121p1101A}
{Abbott}, B.P.; et al.
\newblock {GW170817: Measurements of Neutron Star Radii and Equation of State}.
\newblock { Phys. Rev. Lett.} {\bf 2018}, {\em 121},~161101.

\bibitem[{De} \em{et~al.}(2018){De}, {Finstad}, {Lattimer}, {Brown}, {Berger},
  and {Biwer}]{2018PhRvL.121i1102D}
{De}, S.; {Finstad}, D.; {Lattimer}, J.M.; {Brown}, D.A.; {Berger}, E.;
  {Biwer}, C.M.
\newblock {Tidal Deformabilities and Radii of Neutron Stars from the
  Observation of GW170817}.
\newblock { Phys. Rev. Lett.} {\bf 2018}, {\em 121},~091102.

\bibitem[{Yu} and {Weinberg}(2017)]{2017MNRAS.464.2622Y}
{Yu}, H.; {Weinberg}, N.N.
\newblock {Resonant tidal excitation of superfluid neutron stars in coalescing
  binaries}.
\newblock {MNRAS} {\bf 2017}, {\em 464},~2622.

\bibitem[{Lai}(1994)]{1994MNRAS.270..611L}
{Lai}, D.
\newblock {Resonant oscillations and tidal heating in coalescing binary neutron
  stars}.
\newblock {MNRAS} {\bf 1994}, {\em 270},~611.

\bibitem[{Reisenegger} and {Goldreich}(1994)]{1994ApJ...426..688R}
{Reisenegger}, A.; {Goldreich}, P.
\newblock {Excitation of neutron star normal modes during binary inspiral}.
\newblock {Ap. J.} {\bf 1994}, {\em 426},~688.

\bibitem[{Kokkotas} and {Sch\"afer}(1995)]{1995MNRAS.275..301K}
{Kokkotas}, K.D.; {Sch\"afer}, G.
\newblock {Tidal and tidal-resonant effects in coalescing binaries}.
\newblock {MNRAS} {\bf 1995}, {\em 275},~301.

\bibitem[{Ho} and {Lai}(1999)]{1999MNRAS.308..153H}
{Ho}, W.C.G.; {Lai}, D.
\newblock {Resonant tidal excitations of rotating neutron stars in coalescing
  binaries}.
\newblock {MNRAS} {\bf 1999}, {\em 308},~153.

\bibitem[{Reisenegger}(1994)]{1994ApJ...432..296R}
{Reisenegger}, A.
\newblock {Multipole Moments of Stellar Oscillation Modes}.
\newblock { Ap. J.} {\bf 1994}, {\em 432},~296.

\bibitem[{Andersson} and {Pnigouras}(2020)]{2020PhRvD.101h3001A}
{Andersson}, N.; {Pnigouras}, P.
\newblock {Exploring the effective tidal deformability of neutron stars}.
\newblock {Phys. Rev. D} {\bf 2020}, {\em 101},~083001.

\bibitem[{Pratten} \em{et~al.}(2020){Pratten}, {Schmidt}, and
  {Hinderer}]{2020NatCo..11.2553P}
{Pratten}, G.; {Schmidt}, P.; {Hinderer}, T.
\newblock {Gravitational-wave asteroseismology with fundamental modes from
  compact binary inspirals}.
\newblock { Nature Communications} {\bf 2020}, {\em 11},~2553.

\bibitem[{Andersson} and {Pnigouras}(2021)]{2021MNRAS.tmp..395A}
{Andersson}, N.; {Pnigouras}, P.
\newblock {The phenomenology of dynamical neutron star tides}. To appear in
\newblock { MNRAS} {\bf 2021}.

\bibitem[{Poisson} and {Will}(2014)]{2014grav.book.....P}
{Poisson}, E.; {Will}, C.M.
\newblock {\em {Gravity}}; Cambridge University Press, Cambridge,  2014.

\bibitem[{Martynov} \em{et~al.}(2016){Martynov} et~al.]{2016PhRvD..93k2004M}
{Martynov}, D.V.; et al.
\newblock {Sensitivity of the Advanced LIGO detectors at the beginning of
  gravitational wave astronomy}.
\newblock {Phys. Rev. D} {\bf 2016}, {\em 93},~112004.

\bibitem[{Chan} \em{et~al.}(2014){Chan}, {Sham}, {Leung}, and
  {Lin}]{2014PhRvD..90l4023C}
{Chan}, T.K.; {Sham}, Y.H.; {Leung}, P.T.; {Lin}, L.M.
\newblock {Multipolar universal relations between f -mode frequency and tidal
  deformability of compact stars}.
\newblock {Phys. Rev. D} {\bf 2014}, {\em 90},~124023.

\bibitem[{Steinhoff} \em{et~al.}(2016){Steinhoff} et~al.]{2016PhRvD..94j4028S}
{Steinhoff}, J.; et al.
\newblock {Dynamical tides in general relativity: Effective action and
  effective-one-body Hamiltonian}.
\newblock { Phys. Rev. D} {\bf 2016}, {\em 94},~104028.

\bibitem[{Hinderer} \em{et~al.}(2016){Hinderer} et~al.]{2016PhRvL.116r1101H}
{Hinderer}, T.; et al.
\newblock {Effects of neutron-star dynamic tides on gravitational waveforms
  within the effective-one-body approach}.
\newblock { Phys. Rev. Lett.} {\bf 2016}, {\em 116},~181101.

\bibitem[{Foucart} \em{et~al.}(2019){Foucart}, {Duez}, {Hinderer}, {Caro},
  {Williamson}, {Boyle}, {Buonanno}, {Haas}, {Hemberger}, {Kidder}, {Pfeiffer},
  and {Scheel}]{2019PhRvD..99d4008F}
{Foucart}, F.; et al.
\newblock {Gravitational waveforms from spectral Einstein code simulations:
  Neutron star-neutron star and low-mass black hole-neutron star binaries}.
\newblock {Phys. Rev. D} {\bf 2019}, {\em 99},~044008.

\bibitem[{Bauswein} \em{et~al.}(2014){Bauswein}, {Stergioulas}, and
  {Janka}]{2014PhRvD..90b3002B}
{Bauswein}, A.; {Stergioulas}, N.; {Janka}, H.T.
\newblock {Revealing the high-density equation of state through binary neutron
  star mergers}.
\newblock {Phys. Rev. D} {\bf 2014}, {\em 90},~023002.

\bibitem[{Takami} \em{et~al.}(2014){Takami}, {Rezzolla}, and
  {Baiotti}]{2014PhRvL.113i1104T}
{Takami}, K.; {Rezzolla}, L.; {Baiotti}, L.
\newblock {Constraining the equation of state of neutron stars from binary
  mergers}.
\newblock {Phys. Rev. Lett.} {\bf 2014}, {\em 113},~091104.

\bibitem[{Clark} \em{et~al.}(2016){Clark} et~al.]{2016CQGra..33h5003C}
{Clark}, J.A.; others.
\newblock {Observing gravitational waves from the post-merger phase of binary
  neutron star coalescence}.
\newblock {Class. Quantum Grav.} {\bf 2016}, {\em 33},~085003.

\bibitem[{Bauswein} \em{et~al.}(2016){Bauswein}, {Stergioulas}, and
  {Janka}]{2016EPJA...52...56B}
{Bauswein}, A.; {Stergioulas}, N.; {Janka}, H.T.
\newblock {Exploring properties of high-density matter through remnants of
  neutron-star mergers}.
\newblock { European Physical Journal A} {\bf 2016}, {\em 52},~56.

\bibitem[{Bernuzzi} \em{et~al.}(2015){Bernuzzi}, {Dietrich}, and
  {Nagar}]{2015PhRvL.115i1101B}
{Bernuzzi}, S.; {Dietrich}, T.; {Nagar}, A.
\newblock {Modeling the complete gravitational wave spectrum of neutron star
  mergers}.
\newblock {Phys. Rev. Lett.} {\bf 2015}, {\em 115},~091101.

\bibitem[{Chakravarti} and {Andersson}(2020)]{2020MNRAS.497.5480C}
{Chakravarti}, K.; {Andersson}, N.
\newblock {Exploring universality in neutron star mergers}.
\newblock {MNRAS} {\bf 2020}, {\em 497},~5480.

\bibitem[{Spruit} and {Phinney}(1998)]{1998Natur.393..139S}
{Spruit}, H.; {Phinney}, E.S.
\newblock {Birth kicks as the origin of pulsar rotation}.
\newblock { Nature} {\bf 1998}, {\em 393},~139.

\bibitem[{Lockitch} and {Friedman}(1999)]{1999ApJ...521..764L}
{Lockitch}, K.H.; {Friedman}, J.L.
\newblock {Where are the r-modes of isentropic stars?}
\newblock {Ap. J.} {\bf 1999}, {\em 521},~764.

\bibitem[{Berti} \em{et~al.}(2005){Berti}, {White}, {Maniopoulou}, and
  {Bruni}]{2005MNRAS.358..923B}
{Berti}, E.; {White}, F.; {Maniopoulou}, A.; {Bruni}, M.
\newblock {Rotating neutron stars: an invariant comparison of approximate and
  numerical space-time models}.
\newblock {MNRAS} {\bf 2005}, {\em 358},~923.

\bibitem[{Jones} \em{et~al.}(2002){Jones}, {Andersson}, and
  {Stergioulas}]{2002MNRAS.334..933J}
{Jones}, D.I.; {Andersson}, N.; {Stergioulas}, N.
\newblock {Time evolution of the linear perturbations of a rotating Newtonian
  polytrope}.
\newblock {MNRAS} {\bf 2002}, {\em 334},~933.

\bibitem[{Gaertig} and {Kokkotas}(2008)]{2008PhRvD..78f4063G}
{Gaertig}, E.; {Kokkotas}, K.D.
\newblock {Oscillations of rapidly rotating relativistic stars}.
\newblock {Phys. Rev. D} {\bf 2008}, {\em 78},~064063.

\bibitem[Passamonti \em{et~al.}(2009)Passamonti et~al.]{2009MNRAS.394..730P}
Passamonti, A.; others.
\newblock {Oscillations of rapidly rotating stratified neutron stars}.
\newblock {MNRAS} {\bf 2009}, {\em 394},~730.

\bibitem[{Gaertig} and {Kokkotas}(2009)]{2009PhRvD..80f4026G}
{Gaertig}, E.; {Kokkotas}, K.D.
\newblock {Relativistic g-modes in rapidly rotating neutron stars}.
\newblock {Phys. Rev. D} {\bf 2009}, {\em 80},~064026.

\bibitem[{Gaertig} \em{et~al.}(2011){Gaertig} et~al.]{2011PhRvL.107j1102G}
{Gaertig}, E.; others.
\newblock {f-mode instability in relativistic neutron stars}.
\newblock { Phys. Rev. Lett.} {\bf 2011}, {\em 107},~101102.

\bibitem[{Chandrasekhar}(1970)]{1970ApJ...161..561C}
{Chandrasekhar}, S.
\newblock {The effect of gravitational radiation on the secular stability of
  the Maclaurin spheroid}.
\newblock {Ap. J.} {\bf 1970}, {\em 161},~561.

\bibitem[{Schenk} \em{et~al.}(2002){Schenk} et~al.]{2002PhRvD..65b4001S}
{Schenk}, A.K.; others.
\newblock {Nonlinear mode coupling in rotating stars and the r-mode instability
  in neutron stars}.
\newblock {Phys. Rev. D} {\bf 2002}, {\em 65},~024001.

\bibitem[{Arras} \em{et~al.}(2003){Arras} et~al.]{2003ApJ...591.1129A}
{Arras}, P.; others.
\newblock {Saturation of the r-mode instability}.
\newblock {Ap. J.} {\bf 2003}, {\em 591},~1129.

\bibitem[{Pnigouras} and {Kokkotas}(2015)]{2015PhRvD..92h4018P}
{Pnigouras}, P.; {Kokkotas}, K.D.
\newblock {Saturation of the f-mode instability in neutron stars: Theoretical
  framework}.
\newblock {Phys. Rev. D} {\bf 2015}, {\em 92},~084018.

\bibitem[{Ipser} and {Lindblom}(1991)]{1991ApJ...373..213I}
{Ipser}, J.R.; {Lindblom}, L.
\newblock {The oscillations of rapidly rotating Newtonian stellar models. II.
  Dissipative effects}.
\newblock { Ap. J.} {\bf 1991}, {\em 373},~213.

\bibitem[{Lindblom} and {Mendell}(1995)]{1995ApJ...444..804L}
{Lindblom}, L.; {Mendell}, G.
\newblock {Does gravitational radiation limit the angular velocities of
  superfluid neutron stars}.
\newblock {Ap. J.} {\bf 1995}, {\em 444},~804.

\bibitem[{Stergioulas} and {Friedman}(1998)]{1998ApJ...492..301S}
{Stergioulas}, N.; {Friedman}, J.L.
\newblock {Nonaxisymmetric neutral modes in rotating relativistic stars}.
\newblock {Ap. J.} {\bf 1998}, {\em 492},~301.

\bibitem[{Koranda} \em{et~al.}(1997){Koranda}, {Stergioulas}, and
  {Friedman}]{1997ApJ...488..799K}
{Koranda}, S.; {Stergioulas}, N.; {Friedman}, J.L.
\newblock {Upper limits set by causality on the rotation and mass of uniformly
  rotating relativistic stars}.
\newblock { Ap. J.} {\bf 1997}, {\em 488},~799.

\bibitem[{Doneva} \em{et~al.}(2013){Doneva} et~al.]{2013PhRvD..88d4052D}
{Doneva}, D.D.; others.
\newblock {Gravitational-wave asteroseismology of fast rotating neutron stars
  with realistic equations of state}.
\newblock {Phys. Rev. D} {\bf 2013}, {\em 88},~044052.

\bibitem[{Kr{\"u}ger} and {Kokkotas}(2020)]{2020PhRvL.125k1106K}
{Kr{\"u}ger}, C.J.; {Kokkotas}, K.D.
\newblock {Fast Rotating Relativistic Stars: Spectra and Stability without
  Approximation}.
\newblock {Phys. Rev. Lett.} {\bf 2020}, {\em 125},~111106.

\bibitem[{New} \em{et~al.}(2000){New}, {Centrella}, and
  {Tohline}]{2000PhRvD..62f4019N}
{New}, K.C.B.; {Centrella}, J.M.; {Tohline}, J.E.
\newblock {Gravitational waves from long-duration simulations of the dynamical
  bar instability}.
\newblock { Phys. Rev. D} {\bf 2000}, {\em 62},~064019.

\bibitem[{Shibata} \em{et~al.}(2000){Shibata}, {Baumgarte}, and
  {Shapiro}]{2000ApJ...542..453S}
{Shibata}, M.; {Baumgarte}, T.W.; {Shapiro}, S.L.
\newblock {The bar-mode instability in differentially rotating neutron stars:
  Simulations in full general relativity}.
\newblock { Ap. J.} {\bf 2000}, {\em 542},~453.

\bibitem[Baiotti \em{et~al.}(2007)Baiotti et~al.]{2007PhRvD..75d4023B}
Baiotti, L.; others.
\newblock {Accurate simulations of the dynamical bar-mode instability in full
  general relativity}.
\newblock { Phys. Rev. D} {\bf 2007}, {\em 75},~044023.

\bibitem[{Chandrasekhar}(1973)]{1973efe..book.....C}
{Chandrasekhar}, S.
\newblock {\em {Ellipsoidal figures of equilibrium.}}; Dover publications,
  1973.

\bibitem[{Thompson} and {Duncan}(1995)]{1995MNRAS.275..255T}
{Thompson}, C.; {Duncan}, R.C.
\newblock {The soft gamma repeaters as very strongly magnetized neutron stars.
  I. Radiative mechanism for outbursts}.
\newblock {MNRAS} {\bf 1995}, {\em 275},~255.

\bibitem[{Doneva} \em{et~al.}(2015){Doneva}, {Kokkotas}, and
  {Pnigouras}]{2015PhRvD..92j4040D}
{Doneva}, D.; {Kokkotas}, K.; {Pnigouras}, P.
\newblock {Gravitational wave afterglow in binary neutron star mergers}.
\newblock {Phys. Rev. D} {\bf 2015}, {\em 92},~104040.

\bibitem[{Maggiore} \em{et~al.}(2020){Maggiore}, {Van Den Broeck}, {Bartolo},
  {Belgacem}, {Bertacca}, {Bizouard}, {Branchesi}, {Clesse}, {Foffa},
  {Garc{\'\i}a-Bellido}, {Grimm}, {Harms}, {Hinderer}, {Matarrese}, {Palomba},
  {Peloso}, {Ricciardone}, and {Sakellariadou}]{2020JCAP...03..050M}
{Maggiore}, M.; et al.
\newblock {Science case for the Einstein telescope}.
\newblock { JCAP} {\bf 2020}, {\em 2020},~050.

\bibitem[{Kr{\"u}ger} \em{et~al.}(2015){Kr{\"u}ger}, {Ho}, and
  {Andersson}]{2015PhRvD..92f3009K}
{Kr{\"u}ger}, C.J.; {Ho}, W.C.G.; {Andersson}, N.
\newblock {Seismology of adolescent neutron stars: Accounting for thermal
  effects and crust elasticity}.
\newblock {Phys. Rev. D} {\bf 2015}, {\em 92},~063009.

\bibitem[{Andersson} \em{et~al.}(2019){Andersson}, {Haskell}, {Comer}, and
  {Samuelsson}]{2019CQGra..36j5004A}
{Andersson}, N.; {Haskell}, B.; {Comer}, G.L.; {Samuelsson}, L.
\newblock {The dynamics of neutron star crusts: Lagrangian perturbation theory
  for a relativistic superfluid-elastic system}.
\newblock {Class. Quantum Grav.} {\bf 2019}, {\em 36},~105004.

\bibitem[{Samuelsson} and {Andersson}(2009)]{2009CQGra..26o5016S}
{Samuelsson}, L.; {Andersson}, N.
\newblock {Axial quasi-normal modes of neutron stars: accounting for the
  superfluid in the crust}.
\newblock {Class. Quantum Grav.} {\bf 2009}, {\em 26},~155016.

\bibitem[{Gabler} \em{et~al.}(2012){Gabler} et~al.]{2012MNRAS.421.2054G}
{Gabler}, M.; others.
\newblock {Magnetoelastic oscillations of neutron stars with dipolar magnetic
  fields}.
\newblock { MNRAS} {\bf 2012}, {\em 421},~2054.

\bibitem[{Colaiuda} and {Kokkotas}(2012)]{2012MNRAS.423..811C}
{Colaiuda}, A.; {Kokkotas}, K.D.
\newblock {Coupled polar-axial magnetar oscillations}.
\newblock { MNRAS} {\bf 2012}, {\em 423},~811.

\bibitem[{Lander} and {Jones}(2012)]{2012MNRAS.424..482L}
{Lander}, S.K.; {Jones}, D.I.
\newblock {Are there any stable magnetic fields in barotropic stars?}
\newblock { MNRAS} {\bf 2012}, {\em 424},~482.

\bibitem[{Bera} \em{et~al.}(2020){Bera}, {Jones}, and
  {Andersson}]{2020MNRAS.499.2636B}
{Bera}, P.; {Jones}, D.I.; {Andersson}, N.
\newblock {Does elasticity stabilize a magnetic neutron star?}
\newblock { MNRAS} {\bf 2020}, {\em 499},~2636.

\bibitem[{Kr{\"u}ger} \em{et~al.}(2010){Kr{\"u}ger}, {Gaertig}, and
  {Kokkotas}]{2010PhRvD..81h4019K}
{Kr{\"u}ger}, C.; {Gaertig}, E.; {Kokkotas}, K.D.
\newblock {Oscillations and instabilities of fast and differentially rotating
  relativistic stars}.
\newblock { Phys. Rev. D} {\bf 2010}, {\em 81},~084019.

\bibitem[{Lockitch} \em{et~al.}(2001){Lockitch}, {Andersson}, and
  {Friedman}]{2001PhRvD..63b4019L}
{Lockitch}, K.; {Andersson}, N.; {Friedman}, J.
\newblock {Rotational modes of relativistic stars: Analytic results}.
\newblock {Phys. Rev. D} {\bf 2001}, {\em 63},~024019.

\bibitem[{Lockitch} \em{et~al.}(2003){Lockitch}, {Friedman}, and
  {Andersson}]{2003PhRvD..68l4010L}
{Lockitch}, K.; {Friedman}, J.; {Andersson}, N.
\newblock {Rotational modes of relativistic stars: Numerical results}.
\newblock {Phys. Rev. D} {\bf 2003}, {\em 68},~124010.

\bibitem[{Idrisy} \em{et~al.}(2015){Idrisy}, {Owen}, and
  {Jones}]{2015PhRvD..91b4001I}
{Idrisy}, A.; {Owen}, B.J.; {Jones}, D.I.
\newblock {R-mode frequencies of slowly rotating relativistic neutron stars
  with realistic equations of state}.
\newblock { Phys. Rev. D} {\bf 2015}, {\em 91},~024001.

\bibitem[{Centrella} \em{et~al.}(2001){Centrella} et~al.]{2001ApJ...550L.193C}
{Centrella}, J.M.; others.
\newblock {Dynamical rotational instability at low T/W}.
\newblock { Ap. J. Lett.} {\bf 2001}, {\em 550},~L193.

\bibitem[{Watts} \em{et~al.}(2005){Watts}, {Andersson}, and
  {Jones}]{2005ApJ...618L..37W}
{Watts}, A.L.; {Andersson}, N.; {Jones}, D.I.
\newblock {The nature of low $T/|W|$ dynamical instabilities in differentially
  rotating stars}.
\newblock { Ap. J. Lett.} {\bf 2005}, {\em 618},~L37.

\bibitem[{Passamonti} and {Andersson}(2020)]{2020MNRAS.498.5904P}
{Passamonti}, A.; {Andersson}, N.
\newblock {Merger-inspired rotation laws and the low-T/W instability in neutron
  stars}.
\newblock { MNRAS} {\bf 2020}, {\em 498},~5904.

\bibitem[{Xie} \em{et~al.}(2020){Xie}, {Hawke}, {Passamonti}, and
  {Andersson}]{2020PhRvD.102d4040X}
{Xie}, X.; {Hawke}, I.; {Passamonti}, A.; {Andersson}, N.
\newblock {Instabilities in neutron-star postmerger remnants}.
\newblock { Phys. Rev. D} {\bf 2020}, {\em 102},~044040.

\bibitem[{De Pietri} \em{et~al.}(2020){De Pietri}, {Feo}, {Font},
  {L{\"o}ffler}, {Pasquali}, and {Stergioulas}]{2020PhRvD.101f4052D}
{De Pietri}, R.; {Feo}, A.; {Font}, J.A.; {L{\"o}ffler}, F.; {Pasquali}, M.;
  {Stergioulas}, N.
\newblock {Numerical-relativity simulations of long-lived remnants of binary
  neutron star mergers}.
\newblock { Phys. Rev. D} {\bf 2020}, {\em 101},~064052.

\end{thebibliography}
\end{document}